\documentclass[fleqn,usenatbib]{mnras}
\usepackage{newtxtext,newtxmath}

\usepackage[T1]{fontenc}
\usepackage{ae,aecompl}
\usepackage[none]{hyphenat}

\usepackage{scalerel,tikz}
\usetikzlibrary{svg.path}
\definecolor{orcidlogocol}{HTML}{A6CE39}
\tikzset{orcidlogo/.pic={
 \fill[orcidlogocol] svg{M256,128c0,70.7-57.3,128-128,128C57.3,256,0,198.7,0,128C0,57.3,57.3,0,128,0C198.7,0,256,57.3,256,128z};
 \fill[white] svg{M86.3,186.2H70.9V79.1h15.4v48.4V186.2z}
 svg{M108.9,79.1h41.6c39.6,0,57,28.3,57,53.6c0,27.5-21.5,53.6-56.8,53.6h-41.8V79.1z M124.3,172.4h24.5c34.9,0,42.9-26.5,42.9-39.7c0-21.5-13.7-39.7-43.7-39.7h-23.7V172.4z}
 svg{M88.7,56.8c0,5.5-4.5,10.1-10.1,10.1c-5.6,0-10.1-4.6-10.1-10.1c0-5.6,4.5-10.1,10.1-10.1C84.2,46.7,88.7,51.3,88.7,56.8z};
}}
\newcommand\orcidicon[1]{\href{https://orcid.org/#1}{\mbox{\scalerel*{
\begin{tikzpicture}[yscale=-1,transform shape]
\pic{orcidlogo};
\end{tikzpicture}
}{|}}}}

\usepackage{graphicx}	
\usepackage{amsmath}	

\usepackage{gensymb}
\usepackage{pdflscape}
\usepackage{xcolor}
\usepackage{rotating}
\usepackage{lscape}
\usepackage{pdflscape}
\makeatletter
\newcommand\notsotiny{\@setfontsize\notsotiny\@vipt\@viipt}
\makeatother
\usepackage{soul}
\setstcolor{magenta}
\definecolor{Brown}{rgb}{0.647,0.165,0.165}

\title[Faraday Complexity in the Galactic Centre]{Heightened Faraday Complexity in the inner 1 kpc of the Galactic Centre}

\author[Livingston et al.]{
J.\ D.\ Livingston,$^{\orcidicon{0000-0002-4090-8000}\,1}$\thanks{E-mail: jack.david.livingston+academic@gmail.com}
N.\ M.\ McClure-Griffiths,$^{\orcidicon{0000-0003-2730-957X}\,1}$
B. M. Gaensler,$^{\orcidicon{0000-0002-3382-9558}\,2}$
\newauthor
A. Seta$^{\orcidicon{0000-0001-9708-0286}\,1}$
and M. J. Alger,$^{\orcidicon{0000-0001-5110-8845}\,1,3}$
\\  
$^{1}$Research School of Astronomy \& Astrophysics, The Australian National University, Canberra ACT 2611, Australia\\
$^{2}$Dunlap Institute for Astronomy and Astrophysics, 
University of Toronto ON M5S 3H4, Canada\\
$^{3}$Data61, CSIRO, Canberra ACT 2601, Australia\\}

\date{Accepted XXX. Received YYY; in original form ZZZ}

\pubyear{2020}

\begin{document}
\label{firstpage}
\pagerange{\pageref{firstpage}--\pageref{lastpage}}
\maketitle

\begin{abstract}

We have measured the Faraday rotation of 62 extra-galactic background sources in 58 fields using the CSIRO Australia Telescope Compact Array (ATCA) with a frequency range of 1.1 - 3.1 GHz with 2048 channels. Our sources cover a region $\sim 12\, \mathrm{deg}\, \times 12\,\mathrm{deg}$ ($\sim 1 $kpc) around the Galactic Centre region. We show that the Galactic Plane for $|l| < 10^\circ$ exhibits large Rotation Measures (RMs) with a maximum |RM| of $1691.2 \pm 4.9\, \mathrm{rad}\,\mathrm{m}^{-2}$ and a mean $|\mathrm{RM}| = 219 \pm 42\,\mathrm{rad}\,\mathrm{m}^{-2}$.  The RMs decrease in magnitude with increasing projected distance from the Galactic Plane, broadly consistent with previous findings. We find an unusually high fraction (95\%) of the sources show Faraday complexity consistent with multiple Faraday components. We attribute the presences of multiple Faraday rotating screens with widely separated Faraday depths to small-scale turbulent RM structure in the Galactic Centre region. The second order structure function of the RM in the Galactic Centre displays a line with a gradient of zero for angular separations spanning $0.83^\circ - 11^\circ$ ($\sim 120 - 1500$ pc), which is expected for scales larger than the outer scale (or driving scale) of magneto-ionic turbulence. We place an upper limit on any break in the SF gradient of 66'', corresponding to an inferred upper limit to the outer scale of turbulence in the inner 1 kpc of the Galactic Centre of $3$ pc. We propose stellar feedback as the probable driver of this small-scale turbulence. 

\end{abstract}




\section{Introduction}
Magnetic fields play a critical role in the dynamics of spiral galaxies, because the energy density of magnetic fields in the Interstellar Medium (ISM) is comparable to the energy densities of the thermal gas and cosmic rays \citep{2012SSRv..166..293H}. ISM magnetic fields have strengths around a few micro-Gauss and affect star formation \citep{2008MNRAS.385.1820P,2015MNRAS.447.3678B,2019FrASS...6....7K} and the spatio-temporal evolution of the ISM \citep{2017MNRAS.469.4985K}.  

Much is still uncertain about the nature of the magnetic fields that permeate the ISM and in particular those within the centre of our Galaxy. This is because the measurement and the associated interpretation of interstellar magnetic field tracers is difficult \citep{2019Galax...7...45S}. The magnetic field of the Galactic Centre has been studied in a limited capacity, focusing on filaments and Sagittarius A* \citep{2013pss5.book..641B,2018MNRAS.476..235R}. 

A few studies have used narrow-bandwidth radio polarisation measurements of Faraday rotation to derive the magnetic field strength within 1 kpc (projected distance) of the Galactic Centre, finding large and highly variable amounts of Faraday rotation, but these studies have been limited in their frequency range, accuracy, and resolution \citep{2003A&A...401.1023R,2005MNRAS.360.1305R,2008A&A...478..435R,2009ApJ...702.1230T}. Narrow frequency ranges can result in ambiguities in the measurements Faraday rotation which can be resolved by observing over a broad range of frequencies \citep{2011AJ....141..191F,2012MNRAS.421.3300O,2016ApJ...825...59A}. The observation of linearly polarised synchrotron radiation using broadband radio interferometry is one powerful tool for measuring interstellar magnetic fields. Typically this synchrotron radiation comes from background Active Galactic Nuclei (AGN) or foreground objects like Supernovae remnants. In this paper, we aim to use broadband polarimetry from AGN to study turbulence and magnetic fields in the Galactic Centre.

When linearly polarised synchrotron radiation enters a magnetised medium, the polarisation angle, $\chi$, is rotated based on the wavelength, $\lambda$, of the radiation due to the Faraday effect. This rotation is known as Faraday rotation. We define $\chi$ in terms of the Stokes parameters Q and U,
\begin{equation}
    \chi = \frac{1}{2} \tan^{-1} \left(\frac{\mathrm{U}}{\mathrm{Q}}\right).
\end{equation}
Faraday rotation occurs due to thermal electrons and magnetic fields in an ionised plasma and is observed primarily at radio frequencies. The rotation is a function of the initial polarisation angle, $\chi_0$, and the wavelength of the observed radiation, $\lambda_{\mathrm{obs}}$, $\chi (\lambda_{\mathrm{obs}}^2) = \chi_0 + \mathrm{RM}\,\lambda_{\mathrm{obs}}^2$. Here, RM is the rotation measure of the region. 

RM describes the magnitude of Faraday rotation of a single Faraday `screen' along a single line-of-sight, modulated by the line-of-sight thermal electron density, $n_e$ (measured in $\mathrm{cm}^{-3}$),
\begin{equation}
\mathrm{RM} \equiv C \int_{\mathrm{there}}^{\mathrm{here}} n_e\,\textbf{B} \cdot d\,\textbf{r}\, [\mathrm{rad}\,\mathrm{m}^{-2}]. \label{eqn:RM}
\end{equation}
Here, \textbf{B} is the magnetic field strength in micro-Gauss, \textbf{r} and d\textbf{r} are the displacement and incremental displacement along the line-of-sight measured in pc from the source (there) to the observer (here), and C is a conversion constant, $C = 0.812 \, \mathrm{rad}\,\mathrm{m}^{-2}\,\mathrm{pc}^{-1}\, \mathrm{cm}^{3}\,\mu \mathrm{G}^{-1}$. 

We can interpret a Faraday screen as a region of free electrons and magnetic fields in the ISM. This region is assumed to not be turbulent or emit polarised synchrotron radiation internally. More complicated Faraday and depolarisation effects can also occur along the line-of-sight, such as external and internal Faraday dispersion \citep{1966ARA&A...4..245G}. These can be caused by turbulent magneto-ionic environments \citep{1998MNRAS.299..189S}, and as such the picture becomes more complicated and requires a more sophisticated framework to analyse, which we describe in Section \ref{sec:RMsyn}. In reality, regions can have synchrotron-emission or turbulent magneto-ionic environments; in the case of synchrotron-emission this most likely occurs in the AGN source or a foreground object \citep{2015ApJ...815...49A}.

\subsection{RM Synthesis}
\label{sec:RMsyn}
Due to limitations in telescope technology, observations of Faraday rotation only used a few discrete wavelengths and the slope of $\chi$ vs $\lambda^2$ was used to measure the RM of a line-of-sight. When there are multiple Faraday screens across a single telescope resolving element (beam), or external and internal Faraday dispersion, we expect the linear relationship of $d \chi / d \lambda^2$ to break down \citep{2011AJ....141..191F,2012MNRAS.421.3300O}. Such sources are known as complex Faraday sources. \cite{1966MNRAS.133...67B} defined a quantity known as the Faraday depth, $\phi$. This quantity indicates the strength and sign of individual Faraday screens within the beam, and is defined similarly to that of the RM,
\begin{equation}
\phi (\textbf{r}) \equiv C \int_{\mathrm{X}}^{\mathrm{here}} n_e\,\textbf{B} \, \cdot d\,\textbf{r}\, [\mathrm{rad}\,\mathrm{m}^{-2}]. \label{eqn:depth}
\end{equation}
Where \textit{X} is the position in space along the line-of-sight, as $\phi$ is a function of the position along the line-of-sight. The Faraday depth function, F($\phi$) \citep{1966MNRAS.133...67B,2005A&A...441.1217B}, is the complex polarised flux density per unit Faraday depth. In the case where there is no external or internal Faraday dispersion from magnetised regions within the beam, we expect each significant peak in the F($\phi$) to correspond to the RM of a Faraday screen.

RM Synthesis is one of the methods of determining F($\phi$), from the complex polarisation vector, $\mathcal{P}(\lambda^{2})$. This requires polarisation data over many frequency channels. $\mathcal{P}$ is related to Stokes Q and U, the polarisation angle, $\chi$, the polarisation fraction \textit{p}, and the total intensity \textit{I} as,
\begin{equation}
    \mathcal{P} = \mathrm{Q} + i \mathrm{U} = p I e^{2i\chi}.
\end{equation}
This relates to F($\phi$) and a `RM spread function' (RMSF) from the sampling in $\lambda^2-\mathrm{space}$, as,
\begin{equation}
    \Tilde{\mathrm{F}}(\phi) = \mathrm{F}(\phi) * \mathrm{RMSF} = K \int_{-\infty}^{\infty} \Tilde{\mathcal{P}}(\lambda^{2})\, e^{-2 i \phi \lambda^2} d\lambda^2.
\end{equation}
Here `$*$' represents a convolution. $\Tilde{\mathcal{P}}(\lambda^{2})$ is the observed complex polarisation vector, $\Tilde{\mathcal{P}}(\lambda^{2}) = W(\lambda^{2}) \mathcal{P}(\lambda^{2})$. $W(\lambda^{2})$ is called the sampling function which is nonzero at all $\lambda^{2}$ points that were measured. $K$ is defined as,
\begin{equation}
    K=\left(\int_{-\infty}^{\infty} W(\lambda^{2}) d(\lambda^{2})\right)^{-1}.
\end{equation}
The `true' F($\phi$) is convolved with RMSF. This creates the observed $\Tilde{\mathrm{F}}(\phi)$ which is highly dependent on the sampling in $\lambda^2-\mathrm{space}$. After deconvolution, the full-width half maximum (FWHM) of the RMSF controls the resolution of the cleaned F($\phi$), $\delta \phi$, \citep{2005A&A...441.1217B,2019ApJ...871..106D}:
\begin{equation}
    \delta \phi \approx \frac{3.79}{\Delta \lambda^2}.
\end{equation}
Here $\Delta \lambda^2 = \lambda_{\mathrm{max}}^{2} -  \lambda_{\mathrm{min}}^{2}$; $\lambda_{\mathrm{max}}^{2}$ and $\lambda_{\mathrm{min}}^{2}$ are the maximum and minimum observed $\lambda^2$. In this study, we use broadband polarisation data, which allow for more complete sampling of the RMSF than previous narrow-band studies, leading to a better F($\phi$) resolution, a larger maximum measurable Faraday depth, maximum measurable-$\phi$-scale, and sensitivity to faint emission components \citep{2005A&A...441.1217B,2019ApJ...871..106D}.

\subsection{Magneto-ionic Turbulence}
\label{sec:turb}
Magneto-ionic turbulence is due to the fluid (or hydrodynamic) turbulence within a magnetised plasma. The second order structure function (SF) has been used to analytically determine important scales in fluids \citep{1941DoSSR..30..301K} and Magneto-Hydrodynamic (MHD) turbulence \citep{1995ApJ...438..763G}. A RM structure function can be used to find the scales over which the product of the magnetic field ($\textbf{B}$) and electron density ($n_e$) varies. The general SF is defined as,
\begin{equation}
\mathrm{SF}_{f}(\delta \theta) = \langle [ f (\theta) - f (\theta + \delta \theta)]^2\rangle_{\theta}.
\label{eqn:SF}
\end{equation}
In this notation, $\theta$ is the angular separation, \textit{f} is the varying quantity, and $\langle ... \rangle_{\theta}$ indicates the average over $\theta$. The $\mathrm{SF}_{\mathrm{RM}} (\delta \theta)$ is twice the variance in RM on a scale of $\delta \theta$. The $\delta \theta$ at which the structure function changes slope tells us about an important scale of fluctuations. For example, a break in the slope of the RM structure function can be related to the outer scale of turbulence and the slope around this scale indicates how turbulence changes with angular scales. The outer scale is synonymous with the driving scale of turbulence.

To relate the slope `break' scale of the RM structure function to the largest scale of magneto-ionic turbulence, we require the assumption that the largest scale of magnetic-ionic variations is comparable to the outer scale of the fluid turbulence. This assumption is motivated by the fact that the correlation scale of magnetic fields in numerical simulations of driven turbulence is comparable to the outer scale of turbulence \citep{SetaEA2020}. Based on the fluid and MHD turbulence theories \citep{1991RSPSA.434....9K,1995ApJ...438..763G},  we expect a power law SF where smaller scales contribute less to the turbulent energy within a region than larger scales. The physical mechanism causing this is a turbulent cascade, in which energy at larger-scales (maintained due to driving) is transferred to smaller-scales (dissipated due to viscosity).

As an example of using the RM structure function to determine the outer scale of fluid turbulence, \cite{2006AN....327..483H} found a zero gradient with an outer scale on the order of $\sim 10$ pc for the inner Galactic Plane, indicating that the main source of turbulence had to be injected on a scale of $\sim 10$ pc, which they attributed to H{\sc ii} regions. 
\\~\\
In this paper, we present the calculated peak Faraday depths of 62 sources close to the Galactic Centre, along with a measure of dispersion between multiple peaks in the F($\phi$) of each source. With these data we construct a RM structure function. The details of the observational data are described in Section \ref{sec:data}. The data are presented in Section \ref{sec:results} in Table \ref{tbl:fara} along with discussion of trends in the spatial distributions of Faraday depth and comparisons to previous studies of the Galactic Centre. Section \ref{sec:results} also contains analysis of the calculated RM structure function. Section \ref{sec:discuss} contains a discussion of the Faraday complexity of our sources, and the possible location along the line-of-sight where the Faraday rotation occurs. Section \ref{sec:cause} contains a discussion of the possible causes for the inferred small scale magneto-ionic turbulence observed within the physical distance structure function. Our conclusions are given in Section \ref{sec:conclude}. 

\section{Data}
\label{sec:data}
\subsection{Observations and Data Reduction}
The observations for this study were obtained through two separate projects on the Australia Telescope Compact Array (ATCA). The first set of data\footnote{Obtained from the \href{https://atoa.atnf.csiro.au/}{Australia Telescope Online Archive}} were observed for a project to study atomic hydrogen absorption through a hydrogen cloud in the foreground of the Galactic Centre \citep{2018MNRAS.479.1465D}. The observations targeted bright compact continuum sources over the region $|l| \lessapprox 10^{\circ}$, $|b| \lessapprox 10^{\circ}$ of the Galactic Centre. The target sources were selected from the NVSS catalogue \citep{1998AJ....115.1693C} and had integrated 1.4 GHz fluxes $\geq$ 200 mJy and were unresolved in NVSS (diameter $\leq$ 45'').  A total of 47 sources were observed with the ATCA in the 1.5C antenna configuration.  In the 1.5C configuration, five of the six 22 m antennas are distributed along a east-west track with a maximum baseline of 1.5km and one antenna (CA06) is fixed at a distance of 3 km from one end of the track. Antenna CA06, providing baselines up to 6km, was available for these observation and used for our imaging. 

Every field was observed for a total of $\sim 100$ minutes between May and June 2015. Observing runs went for 12 hours, the ATCA primary flux and bandpass calibrator, PKS 1934-638, was observed for 30 minutes at the start and end of the 12-hr observing session. Fields were observed hourly, which gave sufficient \textit{uv} coverage for imaging. Data were obtained for both continuum and atomic hydrogen spectral line using the 1M-0.5k correlator configuration on the Compact Array Broadband Backend \citep{2011MNRAS.416..832W}.  Only the continuum data covering all four linear polarisation products, XX, YY, XY and YX, were used here.  The continuum data cover the frequency range 1.1 - 3.1 GHz with 2048 channels. 

Data reduction was carried out with the {\sc Miriad} software package \citep{1995ASPC...77..433S}. We used PKS 1934-638 for bandpass, amplitude and leakage calibration assuming that PKS 1934-638 is un-polarised \citep{2000MNRAS.319..484R}.  The brightest continuum source observed each day was used for phase calibration (NVSS J172920-234535, J174713-192135, J172836-271236, J175233-223012, J175151-252359, J174713-192135, J175114-323538, J174716-191954). The robust visibility weighting \citep{1995AAS...18711202B} was set to +1.5 with a mean pixel size of 2'', an ideal synthesised beam size of $\sim6$'' by $\sim13$'', and a field size of $1000\times1000$ pixels. 

A second set of observations was conducted in January 2019 specifically for this study to extend the sky coverage of the previous data. Eleven additional targets were selected from the NVSS catalogue \citep{2009ApJ...702.1230T} to fill in regions with limited coverage at $|l| \lessapprox 10^{\circ}$, $|b| \lessapprox 10^{\circ}$ from the original survey. The telescope configuration and observation strategy were the same as the previous run of observations. The 6 km baseline available using ATCA was also included for these observations. The total observation run went for 12 hours, each source was observed for a minimum of 1 minute per hour, with a total minimum and maximum integration time of 11 minutes and 176 minutes, respectively. There was 30 minutes of calibration, observing the bandpass calibrator, PKS 1934-638, and 30 minutes of overhead time. The phase and polarisation calibrator used was PKS 1827-360, which has a flux density of $3.986 \pm 0.096$ Jy at 2100 MHz. The phase calibrator was observed every hour for 2 minutes. 

\begin{figure}
    \centering
    \includegraphics[width=\columnwidth]{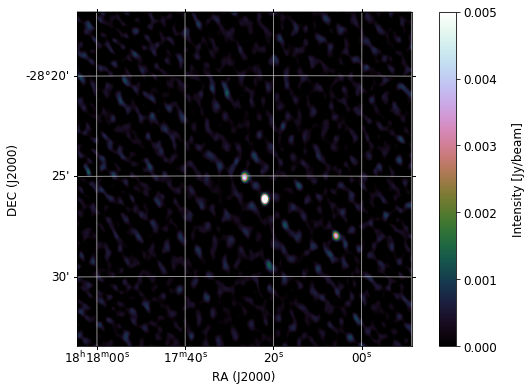}
    \caption{Example Stokes I image of field 1817-2825 (Named based on centre of field) reduced using Miriad software package. The image has; a RMS of $0.001$ Jy/beam, $B_{\mathrm{maj}}=18.6$" and $B_{\mathrm{min}}=7.3$". The colour map is `cubehelix' \protect\citep{2011BASI...39..289G}.}
    \label{fig:example_image}
\end{figure}

Again, data reduction was performed in the software package {\sc Miriad} \citep{1995ASPC...77..433S}. The robust visibility weighting was set to +0.8 with a pixel size of 1'', this robust visibility weighting and pixel size were chosen instead of +1.5 and a pixel size of 2'' as these observations had more complete \textit{uv} coverage than the observations taken in 2015. The field size was of $1000\times1000$ pixels. Data flagging was completed by hand in {\sc Miriad} using the sum-threshold method. After flagging, these data were calibrated and image cubes with 15 MHz channels were created of Stokes parameters I, Q, U, and V, using {\sc invert} and {\sc clean} algorithms with typical rms noise levels of 150 $\mu$Jy/beam. An example Stokes I image of field 1817-2825 is shown in Figure \ref{fig:example_image}.

\subsection{Polarised Data Extraction}
For each field, multiple sources were identified using the  \href{https://github.com/PaulHancock/Aegean/tree/v2.1.0}{Aegean v2.1.0} source finding algorithms \citep{2012MNRAS.422.1812H,2018PASA...35...11H} on each total Stokes I intensity Multi Frequency Synthesis (MFS) image. Data extraction was done by collecting the Stokes I, Q, and U values from spectral cubes at a source position as defined by the generated Aegean catalogue. The RMS of the Stokes I, Q, and U cubes, was found by taking $100 \times 100$ pixel boxes corresponding to $2$'$\,\times\,2$' away from each source position.

If a source had a Signal-to-Noise Ratio (SNR) in Stokes P lower than 3 averaged over all channels, or there were less than 50\% of the original channels of the source after flagging, the data remaining was not used in our analysis and the source was discarded. Channels that contained significant Radio Frequency Interference (RFI) were ignored when collecting the Stokes I, Q, and U data along with the relevant frequencies. The RFI frequencies flagged were broadly consistent for each source and the majority of channels flagged through data reduction were between 1.5 and 1.7 GHz. Major differences in frequency coverage between sources, due to flagging, were due to the location of sources within the field. Sources away from the centre of the field had higher noise and had a larger number of flagged frequencies. 

To test if sources were resolved, we checked the Stokes I profile of each source in R.A. and Decl. against the beam dimensions. If the FWHM of a source in Stokes I was any greater than the beam in either direction with a tolerance of 1'' - 2''\footnote{This tolerance was set to the limit of our resolution; the size of a pixel in degrees.}, that source was designated as resolved. For these sources we re-imaged data cubes using a common beam resolution set to the largest beam associated with the lowest frequency of a cube. Resolved sources are marked with a $\dagger$ in Table \ref{tbl:fara}. 

The Faraday dispersion function, F($\phi$), and RMSF for each source were computed using the {\sc rm synthesis} and {\sc rm clean} \citep{2017ascl.soft08011H} algorithms from the Canadian Initiative for Radio Astronomy Data Analysis (CIRADA) tool-set \href{https://github.com/CIRADA-Tools/RM-Tools/tree/v1.0.1}{RMtools 1D v1.0.1} \citep{2020ascl.soft05003P}. The pipelines find a second order polynomial model for the input Stokes I and subsequently use that to find Stokes Q/I and U/I, using inverse variance weighting. The RMSF for each source was determined using the associated $\lambda^2$ coverage, this resulted in each source having a different RMSF. {\sc rm clean} used a cleaning cutoff set to three times the noise in Stokes Q/I and U/I. The obtained peak Faraday depths and errors are given in Table \ref{tbl:fara}. The RM-synthesis capabilities for this study are shown in Table \ref{tab:faracap}.

\subsection{Faraday Dispersion Function Second Moment}
\label{define:SecondMoment}
\begin{figure*}
    \centering
    \includegraphics[width=1.9\columnwidth]{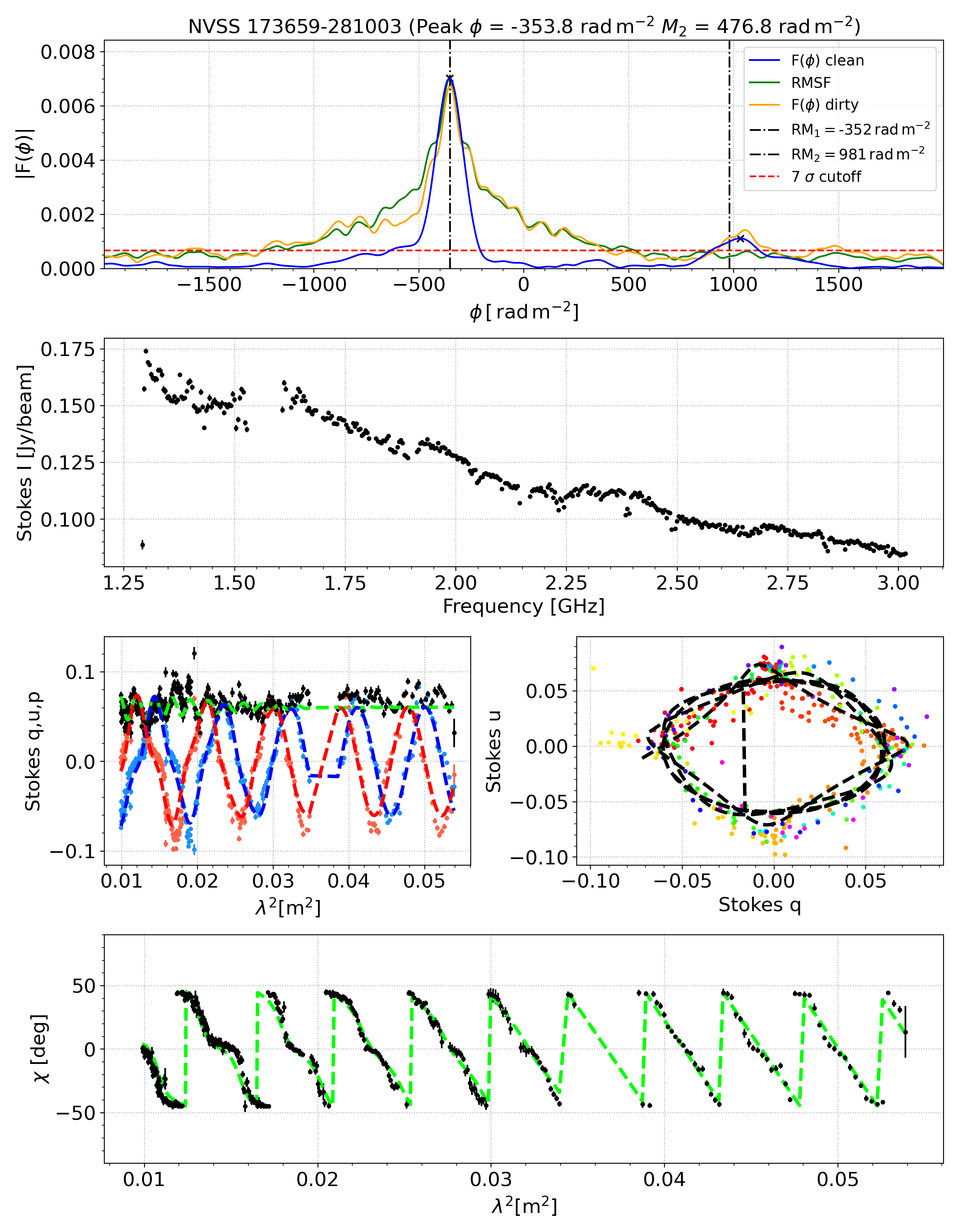}
    \caption{Example of NVSS J173659-281003 that has two highly separated peaks, giving it a large $M_2$. In order from left to right, top to bottom; \emph{Panel 1:} The dirty F($\phi$) is shown in orange, the clean F($\phi$) is shown in blue, the RMSF is shown in green, the peak cutoff line (of seven times the noise) is shown in red, and the black crosses indicate peaks within the spectrum, the black dotted line indicates RMs found through QU Fitting. \emph{Panel 2:} Stokes I against frequency. \emph{Panel 3:} Stokes q (blue), u (red), and p (black) against $\lambda^{2}$ along with model stokes (blue), u (red), and p (green) from QU fitting. \emph{Panel 4:} Stokes q against u, coloured based on $\lambda^{2}$ along with model stokes q against u (black) from QU fitting. \emph{Panel 5:} Polarisation angle, $\chi$, against $\lambda^{2}$ along with model $\chi$ (green) from QU fitting.}
    \label{fig:example_FAR}
\end{figure*}

The majority of our sources (89\%) appear spatially unresolved on the scale of the synthesised beam, which on average was $\approx 26$ by $9$''. But of these 55 unresolved sources, the majority (77\%) show some sub-structure at our maximum baseline of 6km. In addition, for many sources, the F($\phi$) had two or more strong peaks above a SNR of 7 separated in Faraday depth space. An example is shown in Figure \ref{fig:example_FAR}. 

\begin{figure}
    \centering
    \includegraphics[width=\columnwidth]{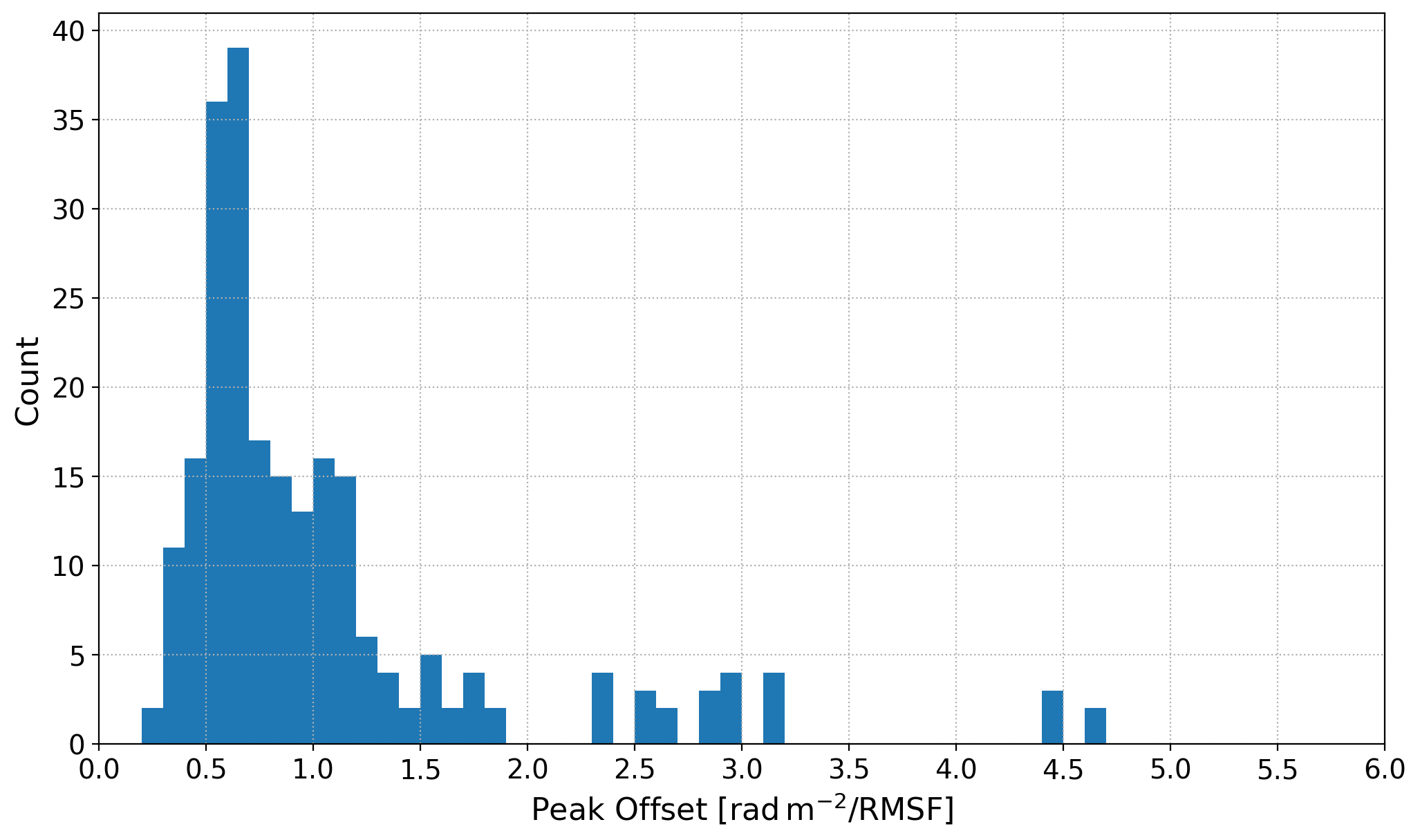}
    \caption{Distribution of Faraday depth peaks above the noise threshold normalised by the difference in $\mathrm{rad}\,\mathrm{m}^{-2}$ between the main peak of the RMSF and the peak of its first sidelobe. Peak offsets that are equal to one could be confused with artefacts from the RMSF.}
    \label{fig:allpeaksoffset}
\end{figure}

A possible interpretation of these multi-component F($\phi$) spectra is that the sources themselves are extended and probe more than one line-of-sight \citep{2005A&A...441.1217B}, but are not resolved by our angular resolution. These separate Faraday depths are most likely to originate in: 1) the AGN themselves or 2) the intervening medium. The other cause of this complexity within a F($\phi$) spectrum may be artefacts from the deconvolution process. 
Faraday depths that are offset from the RMSF peak by exactly the width of the first side-lobe would be indicative of incomplete deconvolution. In Figure \ref{fig:allpeaksoffset} we show the distribution of Faraday depth peaks above the noise threshold normalised with respect to the first RMSF peak (the first side-lobe). As the majority of peak offsets do not correspond to the position of the first RMSF side-lobe we assert the peaks are not artefacts.

We elected to use the second moment, $M_2$, to assess complex F($\phi$) \citep[e.g.][]{2015ApJ...815...49A}. The $M_2$ of a source was derived by masking all peaks in the cleaned F($\phi$) of a source that were less than 7 times the noise as found through RM Synthesis \citep[e.g.][]{2015ApJ...815...49A}. This noise level is a standard cutoff to ensure that the Faraday depths observed are physically real \citep{2012MNRAS.424.2160H,2012ApJ...750..139M}. Faraday depth positions, $\phi_{i}$, were determined using the python \href{https://docs.scipy.org/doc/scipy/reference/generated/scipy.signal.find_peaks.html}{scipy.signal} package. The first moment was calculated as, 
\begin{equation}
    \langle \phi \rangle = J^{-1} \sum_{i=1}^{N} \phi_{i} |\mathrm{F}(\phi_{i})|\,[\,\mathrm{rad}\,\mathrm{m}^{-2}\,],
\end{equation}
$N$ covers all available Faraday depths. Here J, the normalisation constant in units of, is given by,
\begin{equation}
    J = \sum_{i=1}^{N} |\mathrm{F}(\phi_{i})|\,[\,\mathrm{Jy}/\,\mathrm{beam}\,].
\end{equation}
The $M_2$ was calculated as,
\begin{equation}
    M_2 = \sqrt{J^{-1} \sum_{i=1}^{N} (\phi_{i} - \langle \phi \rangle)^{2} |\mathrm{F}(\phi_{i})|}\,[\,\mathrm{rad}\,\mathrm{m}^{-2}\,]. 
    \label{define:FaradayVar}
\end{equation}
The scaling of the separation in Faraday depth from the mean for each source by F($\phi$) amplitude ensures high signal-to-noise peaks are weighted more heavily than lower signal-to-noise peaks. The example spectrum shown in Figure \ref{fig:example_FAR} has a large $M_2 = 476.8\pm6.6\,\mathrm{rad}\,\mathrm{m}^{-2}$. The error in $M_2$ is calculated as the uncertainty in each F($\phi$) spectrum and the standard error in the first moment of F($\phi$). 

Non-zero $M_2$ can derive from: internal Faraday dispersion, $\delta_{\mathrm{RM}}$; external Faraday dispersion $\sigma_{\mathrm{RM}}$, and/or multiple independent Faraday depths, each probing multiple lines-of-sight within the telescope beam, $\langle|\,\mathrm{RM}_1 - \mathrm{RM}_2|\rangle$, \citep{2019MNRAS.487.3432M}. The latter can be observed where the observed AGN is extended, but only on angular scales smaller than the beam. The calculated values of $M_2$ and the associated errors are found in Table \ref{tbl:fara}; all associated F($\phi$) spectra and Stokes information are shown in the supplementary material provided online. 

\section{Results}
\label{sec:results}
We detected 62 polarised sources from the 58 observed fields. Within Table \ref{tbl:fara}, sources were sorted and named based on the closest corresponding NVSS source from \cite{1998AJ....115.1693C}. The mean magnitude RM and median magnitude RM for the surveyed region were $219\pm42$ and $94\pm52$ rad $\mathrm{m}^{-2}$, respectively\footnote{Errors on means and medians were calculated as the standard errors for a sample of 62 sources.}. The standard deviation of |RM| was 329 rad $\mathrm{m}^{-2}$. NVSS J174423-311636 had the greatest magnitude peak Faraday depth of $+1691.2 \pm 4.9 \,\mathrm{rad}\,\mathrm{m}^{-2}$, NVSS J175218-210508 had the smallest magnitude peak Faraday depth of $-3.2 \pm 4.6\, \mathrm{rad}\,\mathrm{m}^{-2}$. For this region, 46\% of the observed sources had positive RMs. 

\begin{figure}
    \centering
    \includegraphics[width=\columnwidth]{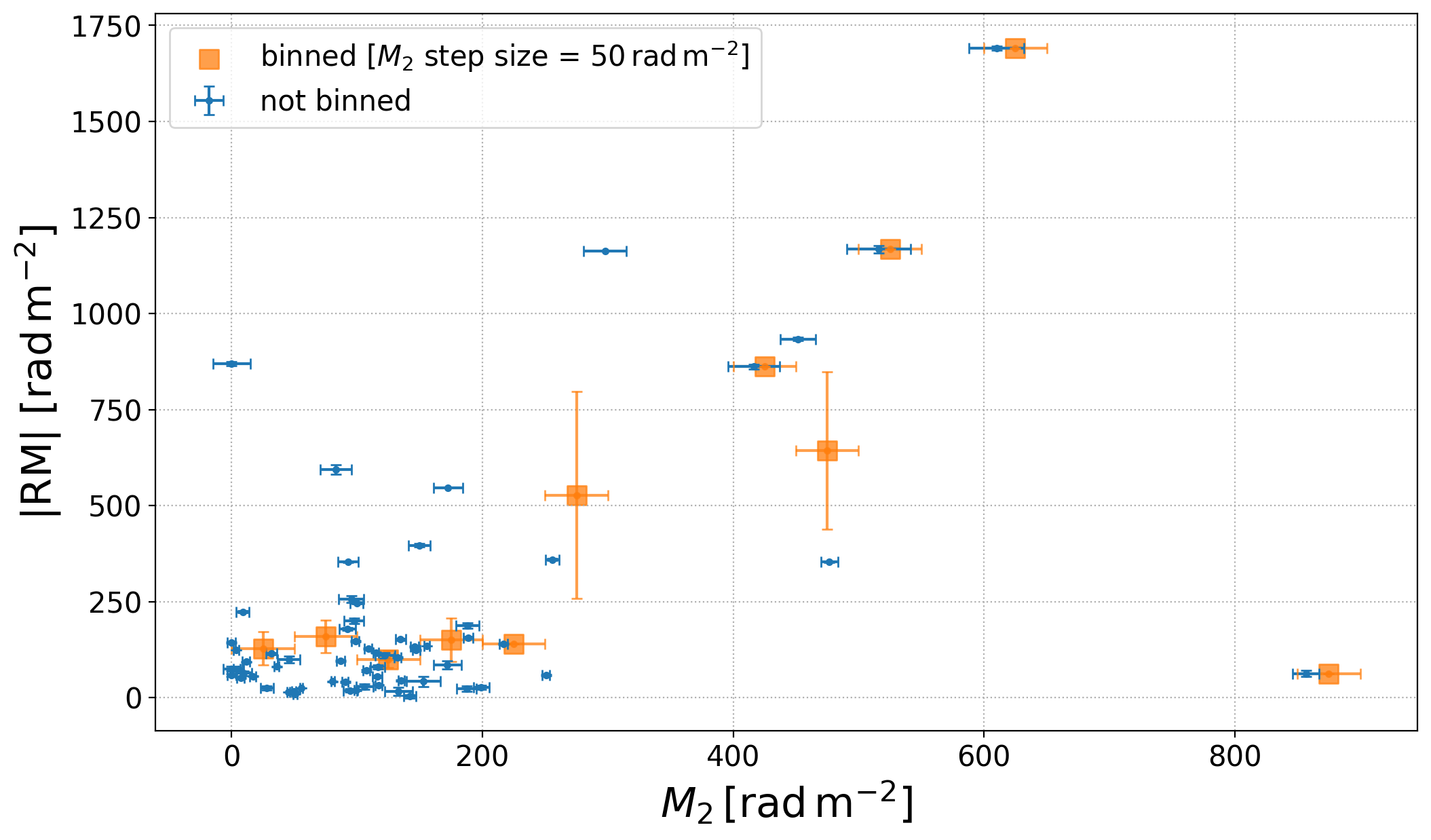}
    \caption{Plot of the absolute value of RM against $M_2$. Binned data is in orange and non-binned data is in blue.}
    \label{fig:m2vsRM}
\end{figure}

For any source with a $M_2$ that is consistent with zero within one standard deviation of its uncertainties, we have assigned $M_2 = 0\,\mathrm{rad}\,\mathrm{m}^{-2}$. Including sources for which $M_2 = 0$, the mean and median $M_2$ were $147 \pm 20\, \mathrm{rad}\,\mathrm{m}^{-2}$ and $103\pm25\, \mathrm{rad}\,\mathrm{m}^{-2}$, respectively. \cite{2015ApJ...815...49A} within their sample found a mean and median $M_2$ of $5.9\pm 1.7\,\mathrm{rad}\,\mathrm{m}^{-2}$ and $0.03 \pm 0.05\,\mathrm{rad}\,\mathrm{m}^{-2}$, respectively. Note that \cite{2015ApJ...815...49A} used a high frequency cutoff of $\sim 2$GHz. The study also found that only 12\% of their sample were `complex' or with a non-zero $M_2$ with a resolution of $\sim 1$'. We also calculated the $M_2$ for all sources found by \cite{2017MNRAS.469.4034O} and found that 55\% of their sources had non-zero $M_2$. We found a mean and median $M_2$ of $5.9\pm1.6\,\mathrm{rad}\,\mathrm{m}^{-2}$ and $0.8\pm 2\,\mathrm{rad}\,\mathrm{m}^{-2}$, respectively and a standard deviation of $16.2\,\mathrm{rad}\,\mathrm{m}^{-2}$. Over the 62 sources, 95\% had a non-zero $M_2$, indicating a high fraction of complex sources in our data compared to previous studies \citep{2015ApJ...815...49A,2017MNRAS.469.4034O}. Our mean $M_2$ is 25 times larger than that of the published mean $M_2$ of both \cite{2015ApJ...815...49A,2017MNRAS.469.4034O}.

The largest values of $M_2$ did not necessarily correspond to the largest peak Faraday depths; the largest $M_2$ was for NVSS J173107-245703 at $856.8 \pm 10.7\, \mathrm{rad}\,\mathrm{m}^{-2}$ and the smallest non-zero $M_2$ was for NVSS J181726-282508-B at $5.5 \pm 4.3\, \mathrm{rad}\,\mathrm{m}^{-2}$. We see no strong correlation between $M_2$ and RM, as shown in Figure \ref{fig:m2vsRM}.

\begin{table*}
\tiny
\centering
\setlength{\tabcolsep}{0.5em}
\begin{tabular}{c c c c c c c c c c @{\vline} c  c c c c c @{\vline} c  c c}
\hline
NVSS (1) & R.A. (2) & Decl. (3) & l (4) & b (5) & Major (6) & Minor (7) & Beam Major (8) & Beam Minor (9) & PI (10) & $\phi$ (11) & (12) & $\mathrm{RM_{tay}}$ (13) & (14)
& $\mathrm{RM_{roy}}$ (15) & (16) & $M_2$ (17) & (18) \\
Name & (J2000) & (J2000) & (deg) & (deg) & (arcsec) & (arcsec) &(arcsec) & (arcsec)& $(\mathrm{Jy/beam})\,\,$ & $\,\,\,(\mathrm{rad}\,\mathrm{m}^{-2})$ & ($\pm$) 
& ($\mathrm{rad}\,\mathrm{m}^{-2}$) & ($\pm$) & $\,\,\,\,$ ($\mathrm{rad}\,\mathrm{m}^{-2}$)$\,\,\,\,$ & ($\pm$)$\,\,\,\,\,\,\,$ 
& $\,\,\,(\mathrm{rad}\,\mathrm{m}^{-2})$ & ($\pm$) \\
\hline
NVSS J172829-284610-A$ \dagger$&17h28m29.63s&-28d46m14.00s&358.13&3.25&24.6&14.1&22.7&8.7&0.7&-79.3&5.1&&&&&116.6&18.2\\
NVSS J172829-284610-B$ \dagger$&17h28m28.87s&-28d46m06.00s&358.13&3.26&30.1&12.1&22.7&8.7&1.1&-75.2&6.4&&&&&0.0&0.0\\
NVSS J172836-271236$ $&17h28m35.85s&-27d12m34.90s&359.44&4.09&24.2&8.9&24.2&8.6&4.0&-70.2&2.4&&&&&107.6&2.8\\
NVSS J172908-265751$ $&17h29m08.09s&-26d57m49.70s&359.72&4.13&24.1&8.8&24.5&8.6&0.7&16.6&10.9&&&&&133.1&11.1\\
NVSS J173107-245703$ $&17h31m06.46s&-24d57m03.90s&1.65&4.86&23.6&9.2&23.9&9.2&0.2&-62.4&8.3&&&&&856.8&10.7\\
NVSS J173133-264015-A$ $&17h31m33.17s&-26d40m15.20s&0.26&3.84&21.8&9.7&24.2&8.6&1.5&-54.7&3.4&&&&&116.1&3.7\\
NVSS J173133-264015-B$ $&17h31m19.00s&-26d38m26.93s&0.25&3.90&29.8&10.4&24.2&8.6&1.2&-58.6&3.2&&&&&0.0&0.0\\
NVSS J173203-285516$ \dagger$&17h32m05.51s&-28d56m06.49s&358.42&2.50&26.7&19.3&20.7&9.2&1.2&-84.3&10.5&&&&&171.9&4.9\\
NVSS J173205-242651-A$ $&17h32m06.05s&-24d26m53.80s&2.20&4.94&29.3&11.1&26.8&8.6&21.9&-14.2&0.3&-3.2&5.7&&&45.5&1.1\\
NVSS J173205-242651-B$ $&17h32m04.73s&-24d26m47.80s&2.19&4.95&29.3&11.1&26.8&8.6&3.9&-9.3&0.8&&&&&50.8&1.5\\
NVSS J173524-251036$ $&17h35m24.91s&-25d10m36.40s&1.99&3.92&26.6&11.7&25.9&8.5&0.5&42.2&13.5&&&&&152.9&13.6\\
NVSS J173659-281003$ $&17h37m00.30s&-28d16m57.63s&359.55&1.95&30.5&12.1&29.7&11.5&7.0&-353.8&2.1&&&&&476.8&6.6\\
NVSS J173713-224734-B$ $&17h37m13.04s&-22d47m34.50s&4.23&4.84&28.9&8.9&28.5&8.6&0.4&-63.5&6.2&&&&&9.6&6.5\\
NVSS J173718-260426$ $&17h37m18.07s&-26d04m24.80s&1.46&3.08&26.9&9.9&22.7&8.9&0.4&-98.7&8.6&&&&&45.7&8.8\\
NVSS J173722-223000$ $&17h37m22.84s&-22d30m00.00s&4.50&4.97&32.6&8.3&32.4&8.1&13.0&41.9&0.5&53.6&6.8&&&80.5&1.4\\
NVSS J173753-254642-A$ $&17h37m53.21s&-25d46m42.20s&1.78&3.12&23.4&9.5&23.1&9.2&0.2&18.7&5.1&&&&&94.4&5.5\\
NVSS J173753-254642-B$ $&17h37m52.17s&-25d46m42.20s&1.78&3.13&23.4&9.5&23.1&9.2&0.2&28.3&6.5&&&&&106.1&6.8\\
NVSS J173811-262441$ \dagger$&17h38m15.29s&-26d26m37.24s&1.26&2.70&24.1&9.0&22.5&9.2&0.2&-23.1&7.0&&&&&187.5&6.1\\
NVSS J173806-262443$ $&17h38m05.91s&-26d24m43.40s&1.27&2.75&23.8&9.9&22.5&9.2&1.9&44.0&2.0&&&&&135.4&2.4\\
NVSS J173850-221918$ $&17h38m50.54s&-22d19m16.30s&4.83&4.78&30.1&12.1&28.7&8.5&1.0&151.8&2.7&&&&&134.8&4.2\\
NVSS J173754-221851$ $&17h38m18.10s&-22d22m40.64s&4.72&4.85&33.6&6.8&28.7&8.5&0.2&187.5&7.7&&&&&188.1&9.4\\
NVSS J174202-271311$ $&17h42m01.90s&-27d13m09.70s&1.05&1.57&30.5&14.7&32.0&14.9&16.4&-134.6&0.6&-100.2&11.9&-136.0&9.0&155.6&2.3\\
NVSS J174224-203729$ $&17h42m25.00s&-20d37m29.10s&6.72&4.96&36.2&10.5&34.5&8.1&5.1&-15.9&0.8&-8.1&10.9&&&49.6&1.4\\
NVSS J174317-305819$ $&17h43m17.88s&-30d58m19.20s&358.00&-0.64&19.4&9.2&19.4&9.2&0.9&933.4&4.4&&&869.0&13.0&451.6&14.0\\
NVSS J174343-182838$ $&17h43m43.33s&-18d28m38.60s&8.72&5.81&35.9&8.3&35.5&8.5&2.2&146.6&2.2&&&&&98.7&3.1\\
NVSS J174411-255208$ $&17h44m32.36s&-25d48m57.62s&2.54&1.83&32.5&11.7&32.3&11.5&3.1&156.5&2.8&&&&&188.7&3.9\\
NVSS J174423-311636$ $&17h44m23.57s&-31d16m36.60s&357.86&-1.00&19.9&8.9&19.4&8.8&0.9&1691.2&4.9&&&1883.0&2.9&609.9&21.9\\
NVSS J174618-193006$ $&17h46m18.05s&-19d30m06.40s&8.16&4.76&35.2&9.0&34.0&8.5&0.8&199.8&6.9&&&&&97.5&7.9\\
NVSS J174712-190550$ $&17h46m37.72s&-18d26m29.90s&9.11&5.24&34.5&8.7&34.5&8.6&0.4&-24.5&4.9&&&&&28.2&5.1\\
NVSS J174716-191954-A*&17h47m31.69s&-19d09m52.22s&9.11&5.24&37.5&10.5&36.6&9.7&25.2&131.2&0.3&109.7&2.9&&&87.1&0.3\\
NVSS J174716-191954-B*$ \dagger$&17h47m12.51s&-19d20m57.33s&9.11&5.24&35.4&8.7&34.7&8.5&65.7&125.3&0.5&&&&&56.3&0.6\\
NVSS J174748-312315$ $&17h47m48.15s&-31d23m27.20s&358.15&-1.68&21.3&19.4&18.1&9.8&2.7&40.0&1.7&&&&&90.6&2.3\\
NVSS J174831-324102-A$ \dagger$&17h48m32.92s&-32d41m06.60s&357.12&-2.48&22.5&9.9&18.6&9.2&4.1&-546.6&1.2&&&&&172.5&10.8\\
NVSS J174831-324102-B$ \dagger$&17h48m30.38s&-32d40m58.60s&357.12&-2.47&21.9&10.0&18.6&9.2&4.7&-351.6&1.0&&&&&93.0&7.6\\
NVSS J174832-225211$ $&17h48m32.37s&-22d52m01.80s&5.53&2.58&29.1&12.5&28.7&8.5&3.9&-104.5&1.4&&&&&132.4&2.6\\
NVSS J174915-200033$ $&17h49m14.88s&-20d00m29.70s&8.08&3.91&39.1&20.1&32.7&8.5&0.2&-110.0&7.4&&&&&122.3&7.8\\
NVSS J174931-210847$ $&17h49m31.57s&-21d08m41.50s&7.13&3.27&31.3&9.5&31.9&8.4&1.7&139.7&1.8&85.2&10.1&&&216.9&3.4\\
NVSS J175104-235215$ $&17h51m04.04s&-23d52m13.20s&4.97&1.57&22.6&10.3&23.4&9.7&0.2&-256.5&8.5&&&&&95.3&9.8\\
NVSS J175114-323538-A$ $&17h51m15.81s&-32d35m54.80s&357.49&-2.93&25.3&12.7&20.6&9.2&11.0&-80.9&0.5&&&&&35.9&2.0\\
NVSS J175114-323538-B$ $&17h51m13.60s&-32d35m32.80s&357.49&-2.92&21.3&10.7&20.6&9.2&25.3&-25.3&0.2&-34.1&3.9&&&55.5&1.1\\
NVSS J175218-210508$ $&17h52m18.30s&-21d05m08.60s&7.51&2.74&28.6&10.0&27.3&8.9&0.8&-3.2&4.6&&&&&142.0&4.9\\
NVSS J175233-223012$ $&17h52m33.17s&-22d30m08.50s&6.32&1.97&29.2&16.0&27.4&8.7&8.7&-19.3&0.9&&&&&99.2&1.5\\
NVSS J175427-235235$ $&17h54m27.38s&-23d52m33.10s&5.36&0.90&26.1&11.7&23.9&9.6&0.7&869.6&5.6&&&851.0&29.0&0.0&0.0\\
NVSS J175423-235205$ $&17h54m13.24s&-23d51m56.84s&5.34&0.95&24.0&9.6&23.9&9.6&0.2&861.7&6.7&&&&&416.5&20.5\\
NVSS J175526-223211$ $&17h55m26.17s&-22d32m11.00s&6.63&1.38&26.7&8.8&27.0&8.8&4.5&-124.1&1.5&&&&&147.2&3.1\\
NVSS J175548-233322$ $&17h55m48.56s&-23d33m22.00s&5.79&0.79&26.2&11.1&24.9&9.5&5.9&1163.3&1.2&&&1144.0&24.0&297.8&17.1\\
NVSS J175533-233259$ $&17h55m39.69s&-23d32m05.87s&5.79&0.83&25.8&11.8&24.9&9.5&0.4&1167.1&9.4&&&&&516.1&25.4\\
NVSS J175727-223901$ $&17h57m27.86s&-22d39m03.90s&6.76&0.92&29.1&14.4&26.5&9.4&0.7&593.4&12.0&&&&&83.1&12.4\\
NVSS J175622-312215$ $&17h56m22.64s&-31d22m16.90s&359.09&-3.24&19.6&8.7&19.2&8.1&6.5&246.5&2.7&253.7&4.4&&&99.7&4.9\\
NVSS J180319-265214$ $&18h03m19.62s&-26d52m12.60s&3.76&-2.33&20.1&10.6&19.6&7.3&1.8&-223.4&2.8&438.3&11.6&&&8.8&5.4\\
NVSS J180316-274810-A$ $&18h03m18.54s&-27d48m19.80s&2.94&-2.79&35.1&22.8&22.4&7.6&14.6&-396.7&5.2&&&&&149.5&8.7\\
NVSS J180316-274810-B$ $&18h03m15.67s&-27d48m04.80s&2.94&-2.77&26.3&10.4&22.4&7.6&17.8&-359.3&2.2&&&&&255.7&5.2\\
NVSS J180356-294716$ $&18h03m56.75s&-29d47m14.70s&1.28&-3.88&21.8&10.4&19.2&7.8&4.5&-95.6&1.4&-139.8&9.0&&&86.8&3.3\\
NVSS J180542-232244$ $&18h05m42.55s&-23d22m46.60s&7.07&-1.09&29.3&7.7&28.1&7.5&8.1&-51.4&1.3&-57.4&1.7&&&7.2&2.8\\
NVSS J180715-230844$ $&18h07m15.12s&-23d08m43.80s&7.45&-1.29&21.8&7.6&21.6&7.2&0.8&178.8&3.7&135.9&17.1&&&92.4&6.6\\
NVSS J180953-302521-A$ $&18h09m53.44s&-30d25m20.30s&1.34&-5.31&17.3&7.4&17.2&7.2&4.6&55.5&0.9&64.7&15.1&&&16.9&2.6\\
NVSS J180953-302521-B$ $&18h09m52.28s&-30d25m20.30s&1.34&-5.31&17.3&7.4&17.2&7.2&1.4&32.7&1.2&&&&&117.1&2.7\\
NVSS J181726-282508-A$ $&18h17m31.20s&-28d24m08.46s&3.92&-5.82&19.6&8.8&18.6&7.3&2.2&58.5&1.2&&&&&250.6&2.8\\
NVSS J181726-282508-B$ $&18h17m26.43s&-28d25m09.50s&3.90&-5.82&19.6&8.8&18.6&7.3&1.2&74.2&3.0&104.3&15.2&&&5.5&4.3\\
NVSS J182057-252813$ $&18h20m57.79s&-25d28m12.00s&6.89&-5.14&22.8&8.3&22.7&8.0&2.1&-26.2&5.4&-273.2&7.1&&&199.2&6.3\\
NVSS J182040-291005$ $&18h20m40.57s&-29d10m04.60s&3.56&-6.79&37.2&9.5&17.6&7.4&1.6&-114.5&3.2&-141.2&10.1&&&31.7&4.4\\
NVSS J182319-272627$ $&18h23m19.80s&-27d26m24.00s&5.38&-6.52&24.5&9.2&23.8&7.7&30.9&93.1&1.2&87.8&2.8&&&11.6&2.9\\

\hline
\end{tabular}
\caption{Results for sources found; with the closest NVSS name, location in ra, dec, Galactic longitude and latitude, Aegean modelled major ($a$) and minor ($b$) axes, and average beam major ($a_{\mathrm{b}}$) and minor ($b_{\mathrm{b}}$) axes, (columns 1 - 9). The table includes the mean polarised intensity percentage (column 10),
 Peak $\phi$, the RM measurement, of a source (columns 11 - 12), the RM measurement from \protect\cite{2009ApJ...702.1230T} and \protect\cite{2005MNRAS.360.1305R} catalogues (columns 13 - 16), and $M_2$ (columns 17 - 18) in $\mathrm{rad}\,\mathrm{m}^{-2}$. Sources with asterisks are the mean of the same source observed on different dates. Sources with $\dagger$ are resolved or partially resolved sources.}
\label{tbl:fara}
\end{table*}

\begin{table}
    \centering
    \scriptsize
    \setlength{\tabcolsep}{0.4em}
    \begin{tabular}{c c c c @{\vline} c c c}
    \hline
         Range (1) & Width (2) & Sensitivity (3) & Resolution (4) $\,\,\,$ &$\,\,\,$ $\delta \phi$ (5) & max-scale (6) & $\phi_{\mathrm{max}}$ (7) \\
         (GHz) & (MHz) & (mJy/beam) & (arcsec) $\,\,\,$ &$\,\,\,$ ($\mathrm{rad}\,\mathrm{m}^{-2}$) &  ($\mathrm{rad}\,\mathrm{m}^{-2}$) &  ($\mathrm{rad}\,\mathrm{m}^{-2}$) \\
         \hline
         1.4 - 3.0 & 4.4 & 7 & $26\times9$ $\,\,\,$&$\,\,\,$ 103 & 317 & 13796 \\
         \hline
    \end{tabular}
    \caption{Table of observational and RM-synthesis capabilities based on frequency range and channel size. Column 1 and 2 are the mean frequency range and channel width of our observations, column 3 is the mean MFS noise over all fields, column 4 is the mean resolution over all fields in major and minor axes. Column 5 is the mean FWHM of RMSF over all sources, column 6 is the mean maximum measurable scale over all sources, and column 7 is the maximum measurable Faraday depth over all sources \protect\citep{2005A&A...441.1217B,2019ApJ...871..106D}.}
    \label{tab:faracap}
\end{table}

\subsection{Comparison to Previous Surveys}
\label{sec:res_compare}
\begin{figure}
    \centering
    \includegraphics[width=\columnwidth]{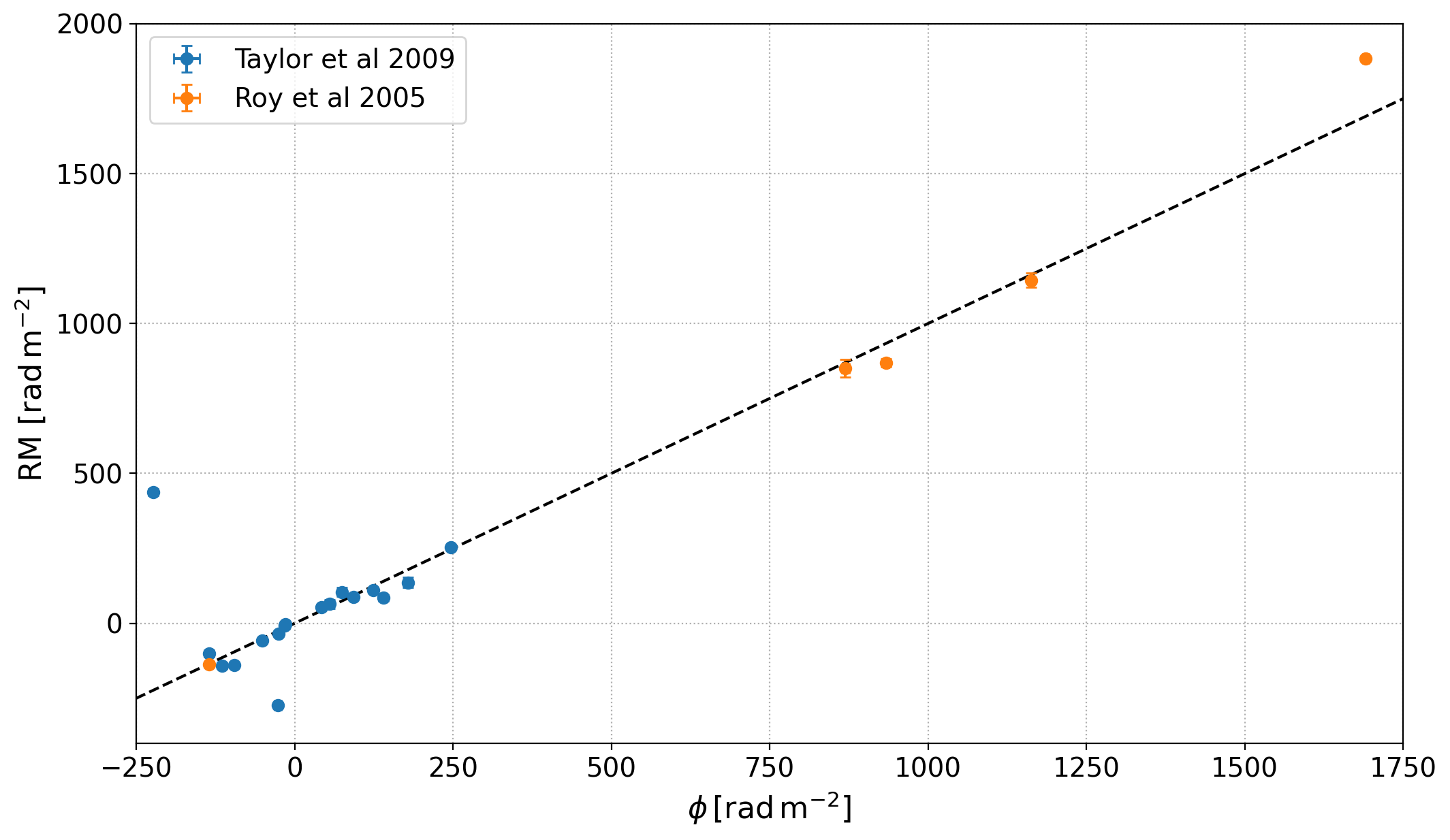}
    \caption{Comparison between our peak Faraday depths and the RMs of \protect\cite{2005MNRAS.360.1305R} and \protect\cite{2009ApJ...702.1230T}.}
    \label{fig:usvstayroy}
\end{figure}

We compared our RMs to published values to check for consistency. \cite{2005MNRAS.360.1305R} and \cite{2009ApJ...702.1230T} both calculated the rotation measures of sources close to the Galactic Centre. We have 17 sources also in \cite{2009ApJ...702.1230T} which were observed using two frequencies, 1364.9 MHz and 1435.1 MHz, each with a width of 42 MHz, and had a beam radius of 45 arcsec. We have 5 sources also observed by \cite{2005MNRAS.360.1305R} observed with 16 discrete frequency bands of 128-MHz from 4.80 to 8.68 GHz and a resolution of $\approx 6$''$\,\times\,2$''. Our mean beam major and minor axes were $\approx 26$ and $9$'', respectively. Both \cite{2005MNRAS.360.1305R} and \cite{2009ApJ...702.1230T} treated all sources as simple rotators with a single RM component; extended sources were split into different sub-sources each with a separate RM measurement. The RMs of the common sources are given in Table \ref{tbl:fara} and a comparison plot is shown in Figure \ref{fig:usvstayroy}. 

The two sources that showed the largest disagreements with the \cite{2009ApJ...702.1230T} catalogue were NVSS J180316-274810-A and NVSS J182057-252813. Our RM for NVSS J180316-274810-A of $-223.5\,\mathrm{rad}\,\mathrm{m}^{-2}$ is within the n$\pi$-ambiguity of 652 rad $\mathrm{m}^{-2}$ discussed by \cite{2019MNRAS.487.3432M,2019MNRAS.487.3454M} for the \cite{2009ApJ...702.1230T} RM catalogue, which could lead to the difference between our RM measurement and that of 438.3 rad $\mathrm{m}^{-2}$ from \cite{2009ApJ...702.1230T}. For the source NVSS J182057-252813, it is likely that there is a confusion between two close objects given that the resolution of \cite{2009ApJ...702.1230T} is larger than the size of our study, which would account for the discrepancies between our observed RM of -26.9 rad $\mathrm{m}^{-2}$ and \cite{2009ApJ...702.1230T} observed RM of -273.2 rad $\mathrm{m}^{-2}$. 

Within our analysis of these sources, we expect our observed peak Faraday depths to be closer to the true peak Faraday depth for each source than those of \cite{2005MNRAS.360.1305R} and \cite{2009ApJ...702.1230T} due to our larger frequency coverage of each source. Faraday depth measurements do not assume a single thin Faraday screen model for each source as is assumed in RM measurements and therefore can extract more information about each source.

\subsection{Spatial Distribution of Measured Values}
\label{sec:spatial}
\begin{figure}
\centering
\includegraphics[width=\columnwidth]{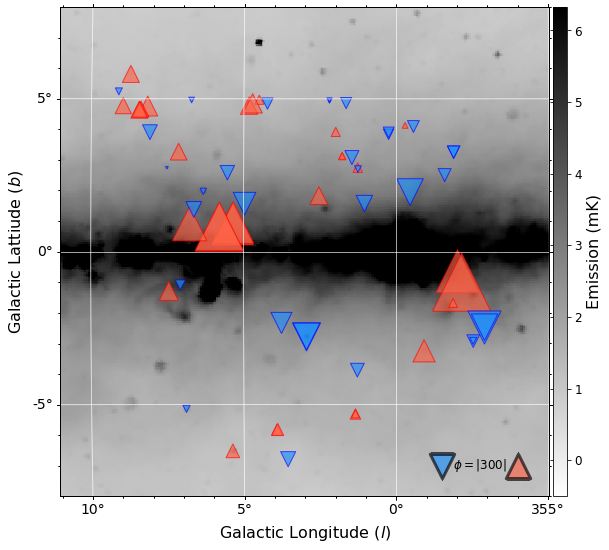}
\caption{Plot of S-PASS 2.3 GHz (grayscale) with peak Faraday depth magnitude markers of all sources over-plotted. The area of each marker corresponds to the magnitude of the measured peak Faraday depth. Negative peak Faraday depths are blue downward triangles, positive peak Faraday depths are red upward triangles.}
\label{fig:Galactic CentrewithRMFD}
\end{figure}

In Figure \ref{fig:Galactic CentrewithRMFD} we show our measured peak Faraday depths plotted over the S-Band Polarization All Sky Survey (S-PASS) 2.3 GHz \citep{2019MNRAS.489.2330C}. S-PASS 2.3 GHz primarily captures synchrotron radiation and thermal emission. There is a general trend of lower magnitude peak Faraday depths farther from the Galactic Centre. We see larger magnitude peak Faraday depths around the Galactic Plane and large positive values in peak Faraday depths around two regions of high continuum intensity at Galactic latitudes and longitudes of l $= 0^{\circ}$, b $= 0^{\circ}$ and l $= 6^{\circ}$, b $= 0^{\circ}$. 

\begin{figure}
\centering
\includegraphics[width=\columnwidth]{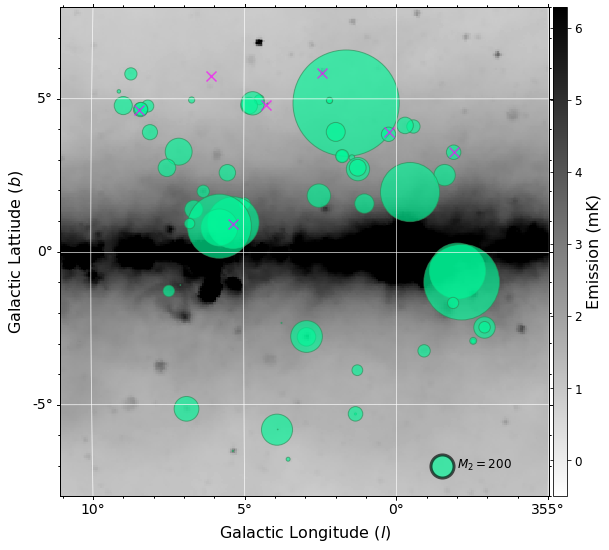}
\caption{Plot of S-PASS 2.3 GHz (grayscale) with peak $M_2$ magnitude markers of all sources over-plotted. The area of each marker corresponds to the square magnitude of $M_2$. Pink crosses indicate sources for which $M_2 = 0$ rad $\mathrm{m}^{-2}$.}
\label{fig:Galactic CentrewithComp}
\end{figure}

\begin{figure}
    \centering
    \includegraphics[width=\columnwidth]{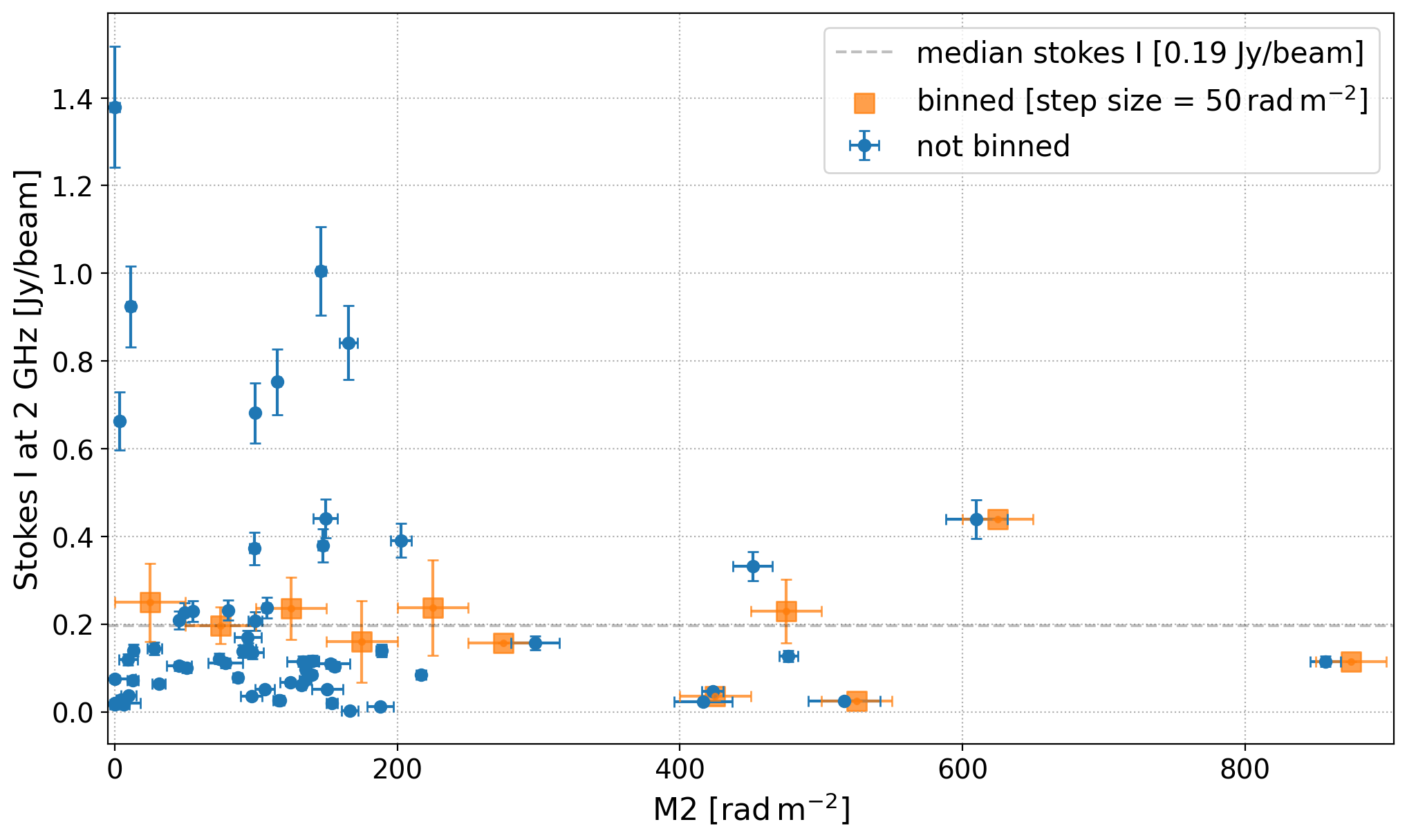}
    \caption{Plot of Stokes I at 2 GHz against $M_2$, binned data is in orange and non-binned data is in blue. The dashed grey line shows the median Stokes I, indicating a lack of correlation between $M_2$ and Stokes I at 2 GHz.}
    \label{fig:m2vsI}
\end{figure}

In Figure \ref{fig:Galactic CentrewithComp} we show our $M_2$ measurements plotted over S-PASS \citep{2019MNRAS.489.2330C} with the area of markers scaled on the squared value of $M_2$. There is a weak trend of larger $M_2$ towards the Galactic Plane. We test for Stokes I leakage from Galactic Centre continuum emission into the polarised emission of each source. We compared the Stokes I at 2 GHz to the $M_2$ for each source shown in Figure \ref{fig:m2vsI} and found no correlation between the two. In our ATCA fields the Stokes I emission is almost exclusively from background AGN. The large-scale emission, like that shown in Figures \ref{fig:Galactic CentrewithRMFD} and \ref{fig:Galactic CentrewithComp}, is resolved out by the extended baselines of the interferometer, leaving only the compact sources. This means it is unlikely that internal Faraday dispersion plays a large role within our sources. Thus, the connection must be between larger $M_2$ and proximity to the Galactic Plane. 


\subsection{Structure Function}
\label{sec:methodSF}
The RM structure function, $\mathrm{SF}_{\mathrm{RM}}$ was calculated by finding the square differences in peak Faraday depth of a source and all other sources within some angular separation (Eq. \ref{eqn:SF}). This was repeated for each source. The resultant differences were binned in 11 bins between $0.83 \degree$ to $11 \degree$. Each bin had at least 20 pairs \citep{2011ApJ...726....4S} and the mean number of pairs per bin was 172. We used the python module \href{bootstrapped v0.0.2}{https://pypi.org/project/bootstrapped/} with a confidence level of 66\% and 10000 iterations to determine the spread for each bin to account for the treatment of errors in log/log plots. This formed the RM structure function for the angular separation between sources. 

The structure function is the sum of various sources of variation \citep{2011ApJ...726....4S},
\begin{equation}
    \mathrm{SF_{RM}} = 2\sigma_{\mathrm{int}}^2 + 2\sigma_{\mathrm{IGM}}^2 + 2\sigma_{\mathrm{ISM}}^2 + 2\sigma_{\mathrm{noise}}^2,
\end{equation}
here $\sigma_{\mathrm{int}}^2$ is the variation of RM generated in the vicinity of the AGN; $\sigma_{\mathrm{IGM}}^2$ is the contribution from the intergalactic medium; $\sigma_{\mathrm{ISM}}^2$ is the contribution of the Galactic Centre and ISM, which is ultimately of the most interest for this study; and $\sigma_{\mathrm{noise}}^2$ is the noise contribution of the uncertainty in measuring the RM of our sources. 

\begin{figure}
\centering
\includegraphics[width=\columnwidth]{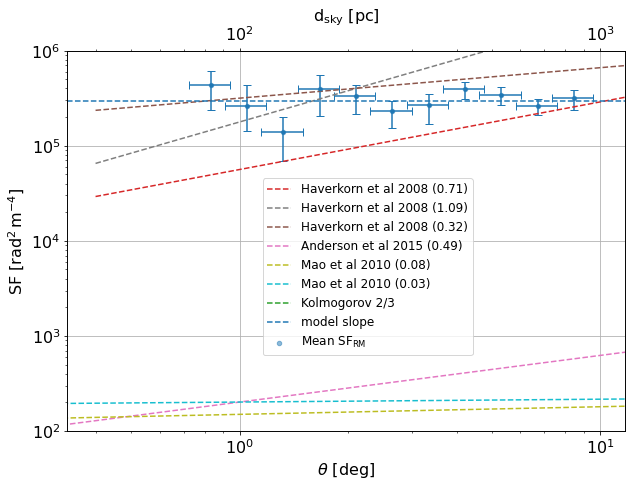}
\caption{Plot of RM structure function as a function of angular separation, $\theta$. The dashed blue line represents the model calculated using our data with a log likelihood minimisation method. The size of the horizontal error bars represent the size in $\theta$ of each bin. The dashed red, grey, brown, and pink lines are the SF models of previous studies of the Galaxy \protect\citep{2008ApJ...680..362H,2010ApJ...714.1170M,2015ApJ...815...49A}. The vertical errors were calculated as described in the main text.}
\label{fig:SF}
\end{figure}

To account for $\sigma_{\mathrm{IGM}}^2$, we subtract twice the intrinsic extra-galactic RM scatter of $36\,\mathrm{rad}^{-2}\,\mathrm{m}^{-4}$ found by \cite{2010MNRAS.409L..99S}. To separate the Galactic Centre and ISM contribution on $\sigma_{\mathrm{ISM}}^2$, we subtract twice the intrinsic Milky Way ISM scatter of RM found by \cite{2010MNRAS.409L..99S} of $64\,\mathrm{rad}^{-2}\,\mathrm{m}^{-4}$. To subtract this contribution we use a stochastic approach choosing a Milky Way contribution with fluctuations between 0 to 7 $\sigma_{\rm MW}$. \cite{2015ApJ...815...49A} study was centred on a region away from the Galactic Plane\footnote{RA = 03h29m40s, DEC = -36d16m30s with a grid spanning $7.5 \degree$ in RA and $5.5 \degree$ in DEC.} with 563 polarised sources. Their sources had a mean and median magnitude RM of 13 $\pm$ 3 rad $\mathrm{m}^{-2}$ and 12 $\pm$ 3 rad $\mathrm{m}^{-2}$, respectively, with a standard deviation of 33 rad $\mathrm{m}^{-2}$. We subtract twice the square of standard deviation in RM calculated by \cite{2015ApJ...815...49A} of 33 rad $\mathrm{m}^{-2}$, to account for the intrinsic scatter of background AGN, $\sigma_{\mathrm{int}}^2$, as they studied a region of the sky away from the Galactic Plane. We follow the same approach as \cite{2004ApJ...609..776H} when determining and accounting for $\sigma_{\mathrm{noise}}^2$. This approach separates the RM structure function from a `noise' structure function which is calculated as a Gaussian with width of $\sqrt{\mathrm{noise}^{2}}$. The resultant structure function is shown in Figure \ref{fig:SF}. 

The slope fit of Figure \ref{fig:SF} was a zero gradient over the entire scale range. The equation describing the zero gradient model was
\begin{equation}
    \log_{10}(\mathrm{SF}_{\mathrm{RM}}) = 5.5 \pm 0.5.
\end{equation} 
This slope was calculated using a log likelihood minimisation method which included the error bounds to determine the fit. A zero gradient indicates that the outer scale of turbulence has not been reached for angular scales between $0.83 \degree$ - $11 \degree$ and must be smaller than $0.83 \degree$. The outer scale is the angular separation at which the structure function `breaks' or changes slope (see Section \ref{sec:turb}). Typically, the turbulence contribution from smaller scales than the outer scale follows a positive power-law slope, as turbulence on smaller scales decay faster than that at large scales. When comparing our structure function to the structure functions of \cite{2008ApJ...680..362H}, \cite{2010ApJ...714.1170M}, and \cite{2015ApJ...815...49A}, as in Figure \ref{fig:SF}, we see that for the probed ranges of those studies the structure function slopes are all steeper than our observed zero gradient. We also note that the structure function amplitude of \cite{2010ApJ...714.1170M,2015ApJ...815...49A} is significantly smaller than the structure function in Figure \ref{fig:SF}.

\section{Discussion}
\label{sec:discuss}
There are several strong indicators of enhanced fluctuations on small angular-scales of the Faraday rotating medium in the direction of the Galactic Centre. First, 95\% of our sources have non-zero $M_2$ measurements, suggesting measurable spatially imposed RM differences on the scale of the telescope beam, which is significant when compared to other studies of $M_2$ \citep[e.g.][]{2015ApJ...815...49A}. Second, the amplitude of the structure function in Figure \ref{fig:SF} is higher than found in other parts of the Galaxy \citep[e.g.][]{2010ApJ...714.1170M,2015ApJ...815...49A}. 

To understand the large non-zero $M_2$ measurements, we need to understand what causes them and where along the line-of-sight they originate. We use QU fitting to test if the large $M_2$ are due to multiple Faraday screens or large $\sigma_{\mathrm{RM}}$. QU fitting can recover properties of multiple emission components and the quantities are physically motivated, unlike traditional RM-synthesis \citep{2015AJ....149...60S}. 

\subsection{QU Fitting}
\label{sec:QU}
The process of determining the magneto-ionic environment of a polarised signal by modelling the polarised emission is called QU Fitting. For modelling we only included sources with a peak Faraday depth amplitude above a signal-to-noise ratio of 50. We have chosen this cutoff as we are modelling with a large number of degrees of freedom and we want to ensure every measurement is well determined. In their study, \cite{2017MNRAS.469.4034O} found that 89\% of their sources were modelled with 1 to 2 screens. They also found that 67\% of sources had external Faraday dispersion, and only 9\% of their sources required internal Faraday dispersion in the best-fit model. As discussed in \ref{sec:spatial} internal Faraday dispersion does not plays a large role within our sources. So, we have elected to use models with either one or two screens with individual external Faraday dispersion components;
\begin{equation}
        \mathcal{P}_{j} = p_j \exp[2i \psi_{0,j} + \mathrm{RM}_{j} \lambda^{2}] \exp[- 2 \sigma_{\mathrm{RM},j} \lambda^{4}]
\end{equation}
Where $\psi_{0,j}$ is the initial polarisation angle of a screen, \textit{j}. Two screens were modelled by the addition of two separate $\mathcal{P}_{j}$ components. The parameters for the best fit model for each source is found in Table \ref{tab:QUfitting}.

We used \href{https://github.com/CIRADA-Tools/RM-Tools/tree/v1.0.1}{RMtools 1D v1.0.1} to model each source and used reduced $\chi^{2}$ to determine the best model. For sources in which the reduced $\chi^{2}$ for 1 and 2 screens models were within 10\% of each other, we elected to model the source with a single screen to avoid over-modelling. The most common model was the two screen model making up 95\% of our sources. This is a higher proportion than the number of sources modelled with two screen in \cite{2017MNRAS.469.4034O} of 52\%. The prevalence of sources with two screens is interesting and could related to the underlying structure of background radio sources; sources that have two prominent lobes might typically probe two Faraday screens which could possibly explain why in the set from \cite{2017MNRAS.469.4034O} 52\% of sources were best modelled using two Faraday screens.

The mean and median difference between RM components, $\langle|\,\mathrm{RM}_1 - \mathrm{RM}_2|\rangle$, for our set was $218\pm 39\,\mathrm{rad}\,\mathrm{m}^{-2}$ and $137\pm 49\,\mathrm{rad}\,\mathrm{m}^{-2}$. The mean and median external Faraday dispersion, $\sigma_{\mathrm{RM}}$ for our set was $29\pm 3\,\mathrm{rad}\,\mathrm{m}^{-2}$ and $28\pm 4\,\mathrm{rad}\,\mathrm{m}^{-2}$. The mean and median difference between all Faraday screens, $\langle|\,\mathrm{RM}_i - \mathrm{RM}_j|\rangle$, (here \textit{i} and \textit{j} are all individual screens) for \cite{2017MNRAS.469.4034O} was $52\pm 7\,\mathrm{rad}\,\mathrm{m}^{-2}$ and $19\pm 9\,\mathrm{rad}\,\mathrm{m}^{-2}$. The mean and median $\sigma_{\mathrm{RM}}$ for \cite{2017MNRAS.469.4034O} was $13\pm 2\,\mathrm{rad}\,\mathrm{m}^{-2}$ and $9.9\pm 2\,\mathrm{rad}\,\mathrm{m}^{-2}$. Our mean $\sigma_{\mathrm{RM}}$ and $\langle|\,\mathrm{RM}_1 - \mathrm{RM}_2|\rangle$ are 2 and 4 times larger than that of the means of  \cite{2017MNRAS.469.4034O}, respectively. We surmise that we are probing a significantly different environment than that of \cite{2017MNRAS.469.4034O} as we expect the background AGN to be of a similar population. From this we infer that the enhanced $M_2$ is likely primarily caused by enhanced $\langle|\,\mathrm{RM}_1 - \mathrm{RM}_2|\rangle$ and modulated slightly by enhanced $\sigma_{\mathrm{RM}}$. This would explain why we don't see a complete agreement between $\langle|\,\mathrm{RM}_1 - \mathrm{RM}_2|\rangle$ and $M_2$. The question still remains where this complex magneto-ionic environment occurs along the line-of-sight. In the following subsection, we demonstrate that the fluctuations can reasonably be attributed to the magneto-ionic environment of the Galactic Centre.


\begin{table*}
\centering
\setlength{\tabcolsep}{0.5em}
    \begin{tabular}{c  c c c c c c c c c c c c}
    \hline
NVSS (1) & $\mathrm{Pol}_{1}$ (2) & $\pm$ (3) & $\mathrm{Pol}_{2}$ (4) & $\pm$ (5) & $\mathrm{RM}_{1}$ (6) & $\pm$ (7) & $\mathrm{RM}_{2}$ (8) & $\pm$ (9) & $\sigma_{\mathrm{RM},1}$ (10) & $\pm$ (11) & $\sigma_{\mathrm{RM},2}$ (12) & $\pm$ (13) \\
Name & (\%) && (\%) && ($\mathrm{rad}\,\mathrm{m}^{-2}$) && ($\mathrm{rad}\,\mathrm{m}^{-2}$) && ($\mathrm{rad}\,\mathrm{m}^{-2}$) && ($\mathrm{rad}\,\mathrm{m}^{-2}$) \\
\hline
NVSS J172836-271236&3.2&0.05&2.3&0.22&-63&0.4&-412&23.8&29&0.4&96&3.0 \\
NVSS J173133-264015-A&1.7&0.05&7.5&0.07&267&1.6&-52&0.2&9&1.1&1&0.5 \\
NVSS J173133-264015-B&8.7&0.02&2.6&0.02&-61&0.2&37&0.7&15&0.1&14&0.4 \\
NVSS J173205-242651-A&13.0&0.00&7.4&0.00&-10&0.0&-41&0.0&13&0.0&58&0.0 \\
NVSS J173205-242651-B&3.3&0.79&4.7&0.02&-74&19.8&-11&0.2&83&6.2&1&0.5 \\
NVSS J173659-281003&6.1&0.01&1.8&0.02&-352&0.1&981&0.9&0&0.1&31&0.4 \\
NVSS J173722-223000&6.5&0.00&0.6&0.00&43&0.0&184&0.2&8&0.0&30&0.0 \\
NVSS J173806-262443&4.2&0.03&1.9&0.05&49&0.3&-235&1.5&29&0.2&42&0.6 \\
NVSS J174202-271311&23.6&0.06&14.3&0.46&-129&0.0&-154&2.7&15&0.0&60&0.6 \\
NVSS J174224-203729&3.1&0.00&3.7&0.01&-22&0.0&-74&0.3&14&0.0&67&0.0 \\
NVSS J174343-182838&8.1&0.06&1.4&0.01&371&2.2&149&0.1&108&0.1&30&0.0 \\
NVSS J174716-191954-A*&2.0&0.24&3.7&0.06&128&0.0&-300&0.1&7&0.1&74&0.2 \\
NVSS J174716-191954-B*&17.8&0.03&&&124&0.1&&&27&0.1&& \\
NVSS J174748-312315&2.3&0.01&2.4&0.13&40&0.1&-276&10.5&8&0.1&87&1.1 \\
NVSS J174831-324102-A&4.6&0.00&1.9&0.01&-547&0.0&-609&0.2&1&0.0&27&0.1 \\
NVSS J174831-324102-B&6.7&0.00&6.2&0.00&-356&0.0&-359&0.1&9&0.0&48&0.0 \\
NVSS J174832-225211&7.0&0.00&2.5&0.02&-106&0.0&348&0.3&4&0.0&50&0.1 \\
NVSS J174931-210847&2.7&0.02&6.0&0.18&142&0.3&41&5.3&11&0.2&70&0.9 \\
NVSS J175114-323538-A&2.2&0.02&11.6&0.00&-206&0.6&-84&0.0&42&0.2&13&0.0 \\
NVSS J175114-323538-B&13.1&0.01&5.5&0.16&-27&0.0&-83&2.7&11&0.0&68&0.8 \\
NVSS J175233-223012&1.9&0.06&4.2&0.01&542&1.7&-21&0.1&85&0.4&36&0.0 \\
NVSS J175526-223211&2.0&0.00&0.6&0.01&-124&0.0&-574&1.1&23&0.0&72&0.3 \\
NVSS J175548-233322&5.6&0.00&1.7&0.00&1173&0.0&1270&0.1&16&0.0&27&0.1 \\
NVSS J175622-312215&2.8&0.00&5.2&0.00&1&0.0&250&0.0&11&0.0&6&0.0 \\
NVSS J180319-265214&7.9&0.00&2.2&0.00&-225&0.0&8&0.1&0&0.0&16&0.1 \\
NVSS J180316-274810-A&7.9&0.00&1.4&0.00&-393&0.0&36&0.0&15&0.0&8&0.1 \\
NVSS J180316-274810-B&6.0&0.00&5.2&0.00&-7&0.0&-352&0.0&59&0.0&24&0.0 \\
NVSS J180356-294716&7.9&0.03&11.0&0.01&47&0.1&-112&0.0&55&0.1&20&0.0 \\
NVSS J180542-232244&18.4&0.01&&&-50&0.0&&&0&0.0&& \\
NVSS J180715-230844&1.5&0.08&3.8&0.03&30&2.0&167&0.4&21&1.4&1&0.5 \\
NVSS J180953-302521-A&9.7&0.01&13.4&0.02&76&0.0&30&0.2&29&0.0&67&0.1 \\
NVSS J180953-302521-B&4.6&0.04&5.9&0.02&127&0.4&1&0.2&29&0.2&13&0.1 \\
NVSS J181726-282508-A&4.0&0.04&10.6&0.03&-29&0.6&70&0.0&38&0.3&10&0.2 \\
NVSS J181726-282508-B&4.6&0.09&13.6&0.07&-6&1.1&57&0.5&5&0.9&2&0.5 \\
NVSS J182057-252813&1.0&0.00&23.5&0.19&-10&0.0&318&0.7&2&0.1&129&0.2 \\
NVSS J182040-291005&2.9&0.01&4.6&0.00&36&0.0&-146&0.0&1&0.1&21&0.0 \\
NVSS J182319-272627&12.5&0.01&28.9&0.08&81&0.1&-7&1.2&28&0.0&98&0.0 \\
\hline
    \end{tabular}
    \caption{Results for QU fitting; with the closest NVSS name (column 1). The predicted polarisation fraction for screen 1 and 2 (columns 2 - 5). The RM for screen 1 and 2 (columns 4 - 9). The external Faraday dispersion, $\sigma_{\mathrm{RM,i}}$ for screen 1 and 2 (columns 10 - 13). Sources with asterisks are the mean of the same source observed on different dates.}
    \label{tab:QUfitting}
\end{table*}

\subsection{Locality Argument}
\label{sec:local}
The Galactic Centre is known to exhibit large RMs  \citep{2008A&A...478..435R,2009ApJ...702.1230T,2011ApJ...731...36L}. We compared the RM values that we measure in the direction of the Galactic Centre with other measured values in the Galaxy. To get an indication of typical RM values, we compare our RM values to those of the \cite{2009ApJ...702.1230T} and \cite{2015ApJ...815...49A} catalogues. The \cite{2015ApJ...815...49A} study had a mean and median magnitude RM of 13 $\pm$ 3 rad $\mathrm{m}^{-2}$ and 12 $\pm$ 3 rad $\mathrm{m}^{-2}$, respectively, with a standard deviation of 33 rad $\mathrm{m}^{-2}$. The \cite{2009ApJ...702.1230T} study was of the whole northern sky\footnote{In the catalogue there is a gap at b$\,\sim-30$ and l$\,\sim-50$ of size $50\degree$ due to sky coverage. The density of sampling was also reduced towards the Galactic Plane \citep{2007ApJ...663L..21S}.}. We take a subsection of the catalogue on the Galactic Plane but away from the Galactic Centre for $|b|<40\,\mathrm{deg}$ and $50 < l < 70\,\mathrm{deg}$ which contains 5504 polarised sources. We found a mean and median |RM| of 33 and 23 rad $\mathrm{m}^{-2}$ with a standard deviation of 36 rad $\mathrm{m}^{-2}$. The mean and median |RM| of our data are much greater at $219\pm42$ and $94\pm52$ rad $\mathrm{m}^{-2}$, respectively. From this we can conclude that the process that has caused the RMs for our sources is unusual compared to other regions of the Galaxy. It is therefore reasonable to assume that the magnitude of the RMs we observe originates near or in the Galactic Centre environment. 

The AGN populations of \cite{2015ApJ...815...49A} and \cite{2017MNRAS.469.4034O} are unlikely to be different to that of our sample. We can use these studies to understand the contribution of $M_2$ from the AGN themselves to separate the effects from the intervening medium of the Galactic Centre. By comparison to \cite{2015ApJ...815...49A} and \cite{2017MNRAS.469.4034O} with a mean $M_2$ of $5.9\,\mathrm{rad}\,\mathrm{m}^{-2}$ and medians of  $0.03\,\mathrm{rad}\,\mathrm{m}^{-2}$ and $0.8\,\mathrm{rad}\,\mathrm{m}^{-2}$, respectively, our measured mean and median $M_2$ were much larger at $147 \pm 20\, \mathrm{rad}\,\mathrm{m}^{-2}$ and $103\pm25\, \mathrm{rad}\,\mathrm{m}^{-2}$. For our sources we found 95\% of sources had a non-zero $M_2$ measurement, which is much higher than found by \cite{2015ApJ...815...49A} and \cite{2017MNRAS.469.4034O}. This is not an issue of differing signal-to-noise cutoffs as both our $M_2$ measurement and our calculations of \cite{2017MNRAS.469.4034O} used a cutoff of 7 times the noise of F($\phi$), they also had a similar frequency range and as such a similar max-scale in F($\phi$). \cite{2015ApJ...815...49A} used a minimum single-to-noise ratio cutoff of 6 times the noise of F($\phi$). 

When comparing the results of QU fitting against those of \cite{2017MNRAS.469.4034O}, we found that for our set the mean $\sigma_{\mathrm{RM}}$ was double and the mean difference between the RMs of Faraday screens ($\langle|\,\mathrm{RM}_1 - \mathrm{RM}_2|\rangle$) was 4 times larger than that of \cite{2017MNRAS.469.4034O}. The scale of the RM variations within a resolution element for our measurements, as characterised by $\langle|\,\mathrm{RM}_1 - \mathrm{RM}_2|\rangle$, is large when compared to that expected from the RM variations for other regions in the Galaxy (away from the Galactic Centre on the Plane and away from the Galactic Plane) indicating that the variation likely originates in the Galactic Centre. 

\begin{figure*}
    \centering
    \includegraphics[width=2\columnwidth]{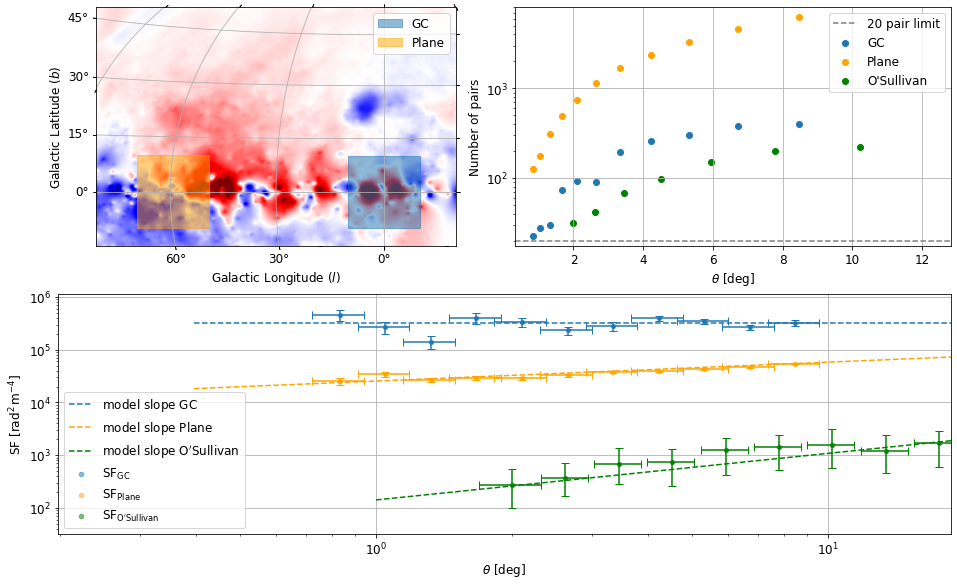}
    \caption{Plot of \protect\cite{2015A&A...575A.118O} RM map with the selected region from the \protect\cite{2009ApJ...702.1230T} catalogue and our surveyed region (top-left). Plot of the angular separation distribution of the Galactic Plane region, the structure function from \protect\cite{2017MNRAS.469.4034O}, and our structure function (top-right), and the resultant structure functions for the region, from \protect\cite{2017MNRAS.469.4034O}, and the structure function from Figure \ref{fig:SF} (bottom). The colour map for the \protect\cite{2015A&A...575A.118O} RM map (top-left) indicates the size and sign of Faraday depth where reds correspond to large positive Faraday depths and blues with large negative Faraday depths; the scale ranges between -500 and 500 rad $\mathrm{m}^{-2}$.}
    \label{fig:pseudoSFs}
\end{figure*}

Another way of comparing the magnitude of our observed RM variations with other areas is with a structure function. We created a structure function of the Galactic Plane from the catalogue of \cite{2009ApJ...702.1230T} and a structure function for the set from \cite{2017MNRAS.469.4034O} to determine if our results shown in Figure \ref{fig:SF} are unusual, compared with other RM catalogues. This was calculated using the same method as described in Section \ref{sec:methodSF}. We used a subsection of the \cite{2009ApJ...702.1230T} catalogue as shown in the top-left of Figure \ref{fig:pseudoSFs} to compute a Galactic Plane RM structure function \citep{2011ApJ...726....4S}. The region selected was a $20\degree \times 20\degree$ box centred on $l = 60\degree,\, b = 0\degree$. The distribution of angular separations for our structure function, the Galactic Plane structure function and the structure function calculated from \cite{2017MNRAS.469.4034O} are shown in the top-right of Figure \ref{fig:pseudoSFs}; the structure functions are presented in the bottom of Figure \ref{fig:pseudoSFs}. 

The approximate slope of the Galactic Plane region SF is,
\begin{equation}
    \log_{10}(\mathrm{SF}_{\mathrm{Plane}}) \sim 1/3 * \log_{10}(\delta \theta) + 4.5.
\end{equation}
The approximate slope of the \cite{2017MNRAS.469.4034O} SF is,
\begin{equation}
    \log_{10}(\mathrm{SF}_{\mathrm{Plane}}) \sim 0.88 * \log_{10}(\delta \theta) + 2.2.
\end{equation}
The value of the Plane region structure function is 1 order of magnitude lower than the height of the structure function shown in Figure \ref{fig:SF}. The height slope of the \cite{2017MNRAS.469.4034O} SF is 3 orders of magnitude lower than the height of our structure function. This significant difference suggests that the amplitude of the flat structure function found in Figure \ref{fig:SF} would not be expected if our surveyed region was away from the Galactic Centre. 

Based on the magnitude of RM fluctuations in the direction of the Galactic Centre and the knowledge that only the Galactic Centre region itself is capable of producing such large RMs, we conclude that the Galactic Centre region is likely responsible for the large $\langle|\,\mathrm{RM}_1 - \mathrm{RM}_2|\rangle$ we observe. If the bulk of the Faraday rotation occurs at the Galactic Centre then these multiple lines-of-sight within a beam are probing the magneto-ionic environment at the distance of the Galactic Centre.

\subsection{Extending the structure function to small scales}
To include small angular scales in the structure function, we use our modelled values of $\langle|\,\mathrm{RM}_1 - \mathrm{RM}_2|^{2}\rangle$ (see Section \ref{sec:QU} and Table \ref{tab:QUfitting}). We note that the RM structure function is defined as the mean square difference of RM across varying angular size scales (shown in Eq. \ref{eqn:SF}). Measuring $\langle|\,\mathrm{RM}_1 - \mathrm{RM}_2|^{2}\rangle$ provides an equivalent extension to scales below the synthesised beam-width. For these scales, we account for the intrinsic RM scatter of multiple screens along the line-of-sight by subtracting the mean squared, $\langle|\,\mathrm{RM}_i - \mathrm{RM}_j|^{2}\rangle$, from \cite{2017MNRAS.469.4034O} of $2700 \pm 700 \,\mathrm{rad}^{2}\,\mathrm{m}^{-4}$ (see Section \ref{sec:QU}). As for the RM structure function (see Section \ref{sec:methodSF}) we subtract the intrinsic scatter of RM for the Milky Way ISM and extra-galactic contribution found by \cite{2010MNRAS.409L..99S} of $64\,\mathrm{rad}^{-2}\,\mathrm{m}^{-4}$ and $36\,\mathrm{rad}^{-2}\,\mathrm{m}^{-4}$, respectively.

\begin{figure}
    \centering
    \includegraphics[width=\columnwidth]{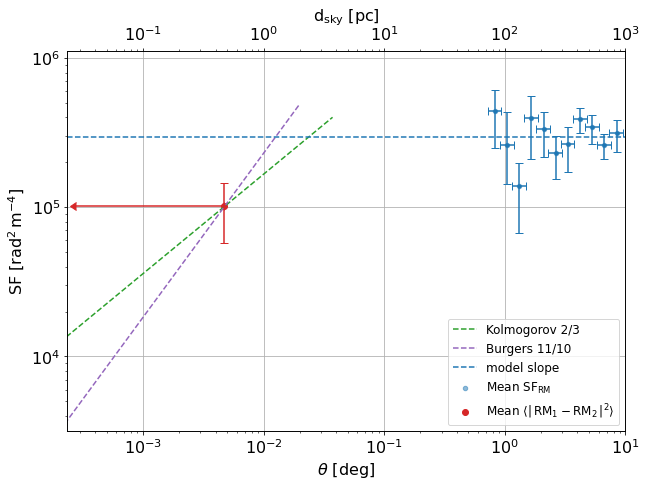}
    \caption{Plot of RM structure function with the mean of $\langle|\,\mathrm{RM}_1 - \mathrm{RM}_2|^{2}\rangle$ as a function of angular separation (bottom axis) and physical distance (top axis). The dashed green and purple lines are the Kolmogorov and Burgers turbulent cascade slopes of 2/3 and 11/10 that intersect with the height of the RM structure function and the mean $\langle|\,\mathrm{RM}_1 - \mathrm{RM}_2|^{2}\rangle$ bin.}
    \label{fig:SFpc}
\end{figure}

For unresolved sources, the angular separation we use the average width of our beams of $17$'', which serves as an upper limit. For resolved sources, we use the angular diameter of the source. The angular separation for the mean $\langle|\,\mathrm{RM}_1 - \mathrm{RM}_2|^{2}\rangle$ bin was taken as the mean of these sizes, or $\sim17$''. This data point serves as an angular separation upper limit as the lines-of-sight within the beam may be closer together than $\sim17$'', but if they are further apart than that they are resolved separately. For reference, the typical sizes of extra-galactic AGN which are between 1" to 30" \citep{1987A&A...179...41O,1993ApJ...405..498W}. We used bootstrap resampling to determine the 66\% confidence interval on the $\langle|\,\mathrm{RM}_1 - \mathrm{RM}_2|^{2}\rangle$ bin. The structure function that includes both the peak Faraday depth and $\langle|\,\mathrm{RM}_1 - \mathrm{RM}_2|^{2}\rangle$ points is shown in Figure \ref{fig:SFpc}.   

We can see from Figure \ref{fig:SFpc}, the $\langle|\,\mathrm{RM}_1 - \mathrm{RM}_2|^{2}\rangle$ point is lower than that of the rest of the structure function. Typically, from the mechanism of a turbulent cascade, we expect smaller angular scales to contribute less to the turbulent energy of a structure function than larger scales. This drop in the turbulent energy contribution at smaller angular scales suggests a possible break in the structure function. 


For sub-sonic (almost incompressible) turbulence we expect turbulent cascade to follow a power-law slope of 2/3 \citep{1991RSPSA.434....9K} below the outer scale. For super sonic (compressible) turbulence we expect a steeper slope of 11/10 \citep{2013MNRAS.436.1245F}. We have plotted both slopes, such that they intersect the height of the $\mathrm{SF}_{\mathrm{RM}}$ and the mean $\langle|\,\mathrm{RM}_1 - \mathrm{RM}_2|^{2}\rangle$ bin in Figure \ref{fig:SFpc}. The gas speed for the ISM over kpc scales is typically trans sonic \citep{2011Natur.478..214G}. As such, we take the mean between the assumptions of sub and super sonic cascade (slopes of 2/3 and 11/10). As we are only fitting a single point this provides us with an upper limit for the mean outer angular scale. This upper limit is $66_{-35}^{+107}$". 

The mean outer scale of the slopes of \cite{2008ApJ...680..362H} was $0.2^\circ$ and for \cite{2015ApJ...815...49A} the outer scale was $1.5^\circ$. These angular outer scales are significantly larger than our estimate of $66_{-35}^{+107}$". \cite{2008ApJ...680..362H} covers an area close to the inner Galactic Plane and the structure function of \cite{2015ApJ...815...49A} covers a high latitude area of the Galaxy and had a significantly smaller amplitude, further indicating we are probing a different magneto-ionic environment.  

By assuming the magneto-ionic environment is at approximately the distance of the Galactic Centre (see Section \ref{sec:local}) we can make the important conversion between the angular separations of sources and the angular scale of the telescope resolution, and a physical scale.  We use the distance to the Galactic Centre of 8.122 $\pm$ 0.031 kpc \citep{2018A&A...615L..15G}, this corresponds to a conversion factor between angular separation to physical distance of $\mathrm{d_{sky}} \sim \tan (\theta)\,\times\,8.122\,\mathrm{kpc}$. With this conversion we construct a structure function for physical distances shown in Figure \ref{fig:SFpc}. The impact parameters at which our sight-lines intersect the Faraday rotating material near the Galactic Centre are at projected galactocentric separations ranging from $269$ pc (NVSS J174202-271311) to $1505$ pc (NVSS J174712-190550). Converting from our previous upper limit of $66_{-35}^{+107}$", this suggests an upper limit for the outer scale of turbulence of $3_{-1}^{+4}\,\mathrm{pc}$. 


\section{Small-scale Magneto-ionic Fluctuations}
\label{sec:cause}
We aim to find a possible cause for the implied outer scale upper limit of turbulence for our studied region of the Galactic Centre of $3$ pc. Outer scales of 10s of parsecs have been observed at high Galactic latitudes \citep{1986ApJ...310..160S} and low Galactic latitudes \citep{1986ApJ...310..160S,1992ApJ...386..143C}, in the spiral arm regions of the Milky Way \citep{2006ApJ...637L..33H,2008ApJ...680..362H}, as well as in the Large Magellanic Cloud where outer scales of $90$ pc were observed \citep{2005Sci...307.1610G}.

Structure functions above the Galactic plane have had shallower slopes, whereas structure functions below the Galactic plane have shown steeper slopes \citep{1986ApJ...310..160S,1992ApJ...386..143C}. Structure functions closer to the Galactic Plane \citep{2004mim..proc.....U,2003A&A...403.1045H,2008ApJ...680..362H} show shallow slopes as well as structure functions of the Fornax region \citep{2015ApJ...815...49A}. A comparison between our structure function and the structure functions of \cite{2008ApJ...680..362H} and \cite{2015ApJ...815...49A}, as well as the aforementioned structure functions around the Galaxy, highlights the flatness of the slope and the particularly small upper limit on the outer scale shown in Figure \ref{fig:SFpc}, especially considering the large range of angular and physical scales we are probing. Given the smaller than usual upper limit on the outer scale, we consider astrophysical effects that are unique to the region between $R_g \sim 270$ pc and $R_g \sim 1500$ pc of the Galactic Centre that could produce an upper limit on the turbulence outer scale of $3\,\mathrm{pc}$. 

Obvious sources of energy injection in this region are: those related to the increased stellar feedback near the Galactic Centre and those related to the complex Galactic dynamics. The majority of Galactic star formation occurs within the Central Molecular Zone of the Galactic Centre \citep{2013MNRAS.429..987L}, so one might expect that stellar feedback could contribute significantly to magneto-ionic turbulence. Similarly, the inner $\sim 3$ kpc of the Milky Way contains a strong bar \citep{doi:10.1146/annurev.astro.34.1.645} that drives non-circular motions, which may inject energy into the Galactic Centre region.  We separately consider the effects of stellar feedback and Galactic dynamics on turbulence below.

\subsection{Stellar Feedback}
We first consider whether stellar feedback may set the outer scale in our data. Stellar feedback, both in the form of stellar winds, H{\sc ii} regions, and Supernovae (SNe), is commonly assumed to be the main source of turbulent energy injection in the Galaxy. Supernovae and super-bubbles are the most powerful interactions, with with theoretical characteristic size scales of $\sim 70$ and $\sim 1000$ pc, respectively \citep{1996ApJ...467..280N,2004RvMP...76..125M}. Observations by \cite{2008ApJ...680..362H} found outer scales $\sim 100$ pc to be caused by SNe. However, the $\sim 100$ pc scale of SNe is much larger than an outer scale inferred from our work, and is therefore an unlikely cause. 

The theoretical size scale of H{\sc ii} regions can be found as \citep{1996ApJ...467..280N},
\begin{equation}
    \mathrm{L_{H_{II}}} = \left(\frac{7 c_{s,\mathrm{II}}}{4 c_{s,\mathrm{I}}}\right)^{4/3} R_s.
\end{equation}
Where $c_{s,\mathrm{I}}$ and $c_{s,\mathrm{II}}$ are the sound speeds of the neutral and ionised media respectively, and $R_s$ is the Str\"{o}mgren radius of the H{\sc ii} region \citep{1939ApJ....89..526S}. At a Galactic Centre H{\sc i} gas density of $10^{4}\,\mathrm{cm}^{-3}$ \citep{2017A&A...603A..90K}, $R_s$ will be between 0.1 - 1 pc. This means that the size scale for the H{\sc ii} regions of the Galactic Centre will be on the order of 10s of parsecs. In their work on RM structure in the Galactic Plane, \cite{2006ApJ...637L..33H} found H~{\sc ii} regions to be the most likely cause of the $\sim$ 17 pc outer scale in inter-arm regions, although they also concluded that it is unlikely a wide-spread phenomenon due to the importance supernovae have in turbulence injection. From the observation and theoretical limits of $L_{\mathrm{H}_{\mathrm{II}}}$, we can rule $\mathrm{H}_{\mathrm{II}}$ regions out as the likely cause our upper limit on the outer scale of 3 pc.

Another source of stellar feedback may be outflows and bubbles around young stars.  Although the size scale for outflows and small bubbles is closer to our 3 pc outer scale, they generally lack power. The expected kinetic energy for these objects is $\sim 10^{43} - 10^{44}$  and $\sim 10^{46}$ ergs \citep{1985ARA&A..23..267L,1996ARA&A..34..111B}, respectively, which is significantly lower than the average kinetic energy of supernovae of $10^{51}$ ergs \citep{1998ApJ...500..342B}.  For the turbulence driven by a spherical explosion with energy $E_{\rm sph}$, the driving scale of turbulence $l_{\rm sph}$ is proportional to $E_{\rm sph}^{1/3}$\citep[Eq. 5.2 of][]{Seta2019}. If $E_{\rm sph}$ is a factor of $10^{6}$ smaller for small bubbles as compared to supernova explosions, $l_{\rm sph}$ should be smaller by a factor of $10^{2}$.  This theory suggests that if SNe inject turbulence on scales of 100 pc, the driving scale of turbulence due to small bubbles will roughly be of the order of 1 pc, which is similar to our observed upper limit for the outer scale. 

Observations on the isolated Taurus molecular cloud by \cite{2015ApJS..219...20L} showed that the turbulence could not be explained with outflows alone. However, the star formation rate within the Central Molecular Zone of the Galactic Centre \citep{doi:10.1146/annurev.astro.34.1.645,2013MNRAS.429..987L} is much higher than in Taurus, and therefore the density of objects may be enough to boost the importance of outflows and small bubbles as a source of turbulent energy injection. We tentatively suggest that stellar outflows and bubbles are a likely candidate, given that only they have enough power in the Galactic Centre, explaining why they are not the main candidate for magneto-ionic turbulence in other galactic regions. 

\subsection{Galactic Dynamics}
\label{section:dyn}
An alternative turbulence injection mechanism may be the unique and powerful dynamics of the Galactic Centre. Within the Galactic Centre's bar exists a resonance between the driving torque of the bar and the gases within. This leads to strong shocks within the gas and deviations from circular motion \citep{doi:10.1146/annurev.astro.34.1.645}. This resonance forms in two regions of the Galactic Centre known as the Inner and Outer Lindblad Resonances. The Inner Lindblad Resonance (ILR) extends out to 1 kpc from the Galactic Centre and is formed due to the orbits effectively overtaking the bar periodically on their journey around the Galactic Centre. The region of sky we are observing sits mostly within the ILR, although we have several lines-of-sight that sit towards the edge of this region. \cite{2017MNRAS.465.1621P} find that this radius ($\sim$ 1 kpc) also corresponds to the point at which the rotation curve of the Milky Way peaks.

\cite{2015MNRAS.453..739K} discuss the instabilities in this region, finding that acoustic instabilities \citep{1999ApJ...520..592M} are more important than gravitational instabilities when understanding the turbulence within the inner 1 kpc of the Galactic Centre. In the Galactic Centre, when gas is perturbed non-axisymmetrically, we see an interaction between the bar and the gas that causes turbulence. The theoretical wavelength scale of acoustic instability turbulence ($L_{\mathrm{ins}}$) found by \cite{2015MNRAS.453..739K} increases outwards from the Galactic Centre, from scales of $L_{\mathrm{ins}} \sim 10^{-2}$ pc at $\sim$ 100 pc from the Galactic Centre to $L_{\mathrm{ins}} \sim 10^{-1}$ pc at $\sim$ 400 pc from the Galactic Centre. We expect this relationship between radius and $L_{\mathrm{ins}}$ to continue out until the ILR at $\sim 1$ kpc. This means that acoustic instability could produce an outer scale of 3 pc or less. 

This effect has been observed by \cite{2016ApJ...832..143F} in their comprehensive observation of G0.253+0.016, a molecular cloud in the CMZ. They found the turbulence driving to be primarily solenoidal, which is unlike molecular clouds found within the spiral arms of the Milky Way \citep{2013ApJ...779...50G}. Solenoidal driving is found to be caused by strong shear, from sharp observed velocity gradients. If this applies to other structures within the CMZ and surrounding regions, we would expect the contribution of shearing forces (and as a result acoustic instability) to play a large role in turbulence injection. 

\cite{2020arXiv200210559S} find that the acoustic instability is a spurious result and may not be able to drive turbulence in the interstellar medium, and instead turbulent viscosity contributes to the observed turbulence. Given this we rule out acoustic instability as a potential cause of our observed upper limit for the outer scale. In their report they find the driving scales of turbulent viscosity are on the order of H{\sc ii} regions, and as such likely also do not generate a turbulence injection scale less than 3 pc. We thus suggest that our inferred outer scale of turbulence is a result of enhanced stellar outflows and bubbles.

\section{Conclusion}
\label{sec:conclude}
We collected broadband polarisation data over 1 - 3 GHz of 58 fields within $\sim 12\degree$ of the Galactic Centre using the ATCA. The peak Faraday depth (RM) and F($\phi$) for the 62 detected polarised sources were calculated using RM Synthesis. The RM measurements were found to be in good agreement with previous observations. We used F($\phi$) for our sources to find the second moment, $M_2$, of Faraday depth and thereby determine the Faraday complexity of each source. The majority (95\%) of our sources had non-zero $M_2$, indicating they have complex RM structure on the scale of the beam. This is significantly higher than the findings of previous studies of Faraday complexity \citep{2015ApJ...815...49A,2017MNRAS.469.4034O}. We modelled our sources using QU fitting and found that 95\% of sources were well modelled with two Faraday screens and external Faraday dispersion. 

We combined the peak Faraday depth data and the mean difference between the RMs for each source, $\langle|\,\mathrm{RM}_1 - \mathrm{RM}_2|^{2}\rangle$, to form a second order RM structure function, which covered angular separations between 17'' and 11\degree. Using an assumption of the location of the surveyed magneto-ionic environment, this resulted in a structure function covering physical distance separations between $0.6 \sim 1500$ pc. 
 
The structure function showed a zero gradient between $\sim120$ pc and 1500 pc. This indicated that for a trans-sonic turbulent cascade (mean of Kolmogorov/Burgers slopes), an upper limit on the outer scale of magneto-ionic turbulence of 3 pc for the Galactic Centre. We discussed Galactic dynamical effects and stellar outflows and bubbles as plausible causes of the observed small scale magneto-ionic turbulence in the Galactic centre. We suggested that the magneto-ionic turbulence may be related to stellar outflows and bubbles, as they have size scales similar to the upper limit on the outer scale of 3 pc.

\section*{Acknowledgements}
The Australia Telescope Compact Array is part of the Australia Telescope National Facility which is funded by the Australian Government for operation as a National Facility managed by CSIRO.  We acknowledge the Gomeroi people as the traditional owners of the Observatory site.  We also acknowledge the Ngunnawal and Ngambri people as the traditional owners and ongoing custodians of the land on which the Research School of Astronomy \& Astrophysics is sited at Mt Stromlo. First Nations people were the first astronomers of this land and make up both an important part of the history of astronomy and an integral part of astronomy going forward. 

We thank the anonymous referee for a thorough review of the work. We thank Craig Anderson and Alec Thomson for helpful discussions related to the paper.  This research was supported by the Australian Research Council (ARC) through grant DP160100723. J.D.L and M.J.A were supported by the Australian Government Research Training Program. N.M.G. acknowledges the support of the ARC through Future Fellowship FT150100024. The Dunlap Institute is funded through an endowment established by the David Dunlap family and the University of Toronto. B.M.G. acknowledges the support of the Natural Sciences and Engineering Research Council of Canada (NSERC) through grant RGPIN-2015-05948, and of the Canada Research Chairs program.

\section*{Data Availability}
The data underlying this article were accessed from the CSIRO Australia Telescope National Facility online archive at \href{https://atoa.atnf.csiro.au}{https://atoa.atnf.csiro.au}, under the project codes C3020 and C3259. The derived data generated in this research will be shared on reasonable request to the corresponding author.



\bibliographystyle{mnras}
\interlinepenalty=10000
\bibliography{example} 




\bsp	
\label{lastpage}
\end{document}